 \documentclass[final,5p,times,twocolumn,authoryear]{elsarticle}

\usepackage{amssymb}

\usepackage{natbib}
\usepackage{longtable}
\bibpunct{(}{)}{;}{a}{,}{,}

\usepackage{graphicx,subcaption}
\usepackage{pifont}
\usepackage{ulem}

\usepackage{pstricks}
\usepackage{color}
\graphicspath{ {figures/}, {koala/}, {genoid/} } 

\usepackage{hyperref}
\hypersetup{
    bookmarks=true,         
    unicode=true,           
    pdftoolbar=true,        
    pdfmenubar=true,        
    pdffitwindow=true,      
    pdftitle={Study of (107) Camilla},
    pdfauthor={Pajuelo et al.},
    pdfsubject={Planetary Science},
    pdfkeywords={},         
    pdfnewwindow=true,      
    colorlinks=true,        
    linkcolor=gray,         
    citecolor=blue,         
    filecolor=gray,         
    urlcolor=gray           
}

\newcommand{\numb}[1]{\textcolor{black}{#1}}
\newcommand{\degr}{\ensuremath{^{\circ}}}
\newcommand{\arcsec}{\ensuremath{^{\prime\prime}}}
\newcommand{\mistral}{\texttt{Mistral}}

\newcommand{\genoid}{\texttt{Genoid}}
\newcommand{\iauOne}{\textsl{S/2001 (107) 1}}
\newcommand{\iauTwo}{\textsl{S/2016 (107) 1}}

\newcommand{\SatOne}{S1}
\newcommand{\SatTwo}{S2}

\newcommand{\ObsOne}{80}
\newcommand{\ObsTwo}{11}

\newcommand{\aOne}{1248}
\newcommand{\pOne}{3.71}

\newcommand{\aTwo}{644}
\newcommand{\pTwo}{1.38}
\newcommand{\eTwo}{0.18}

\newcommand{\rmsOne}{7.8}
\newcommand{\rmsTwo}{5.0}

\newcommand{\KOALA}{\texttt{KOALA}}
\newcommand{\rmsLC}{0.03}
\newcommand{\rmsAO}{0.29}
\newcommand{\rmsOcc}{0.35}

\newcommand{\nbLC}{127}

\newcommand{\Diam}{254}
\newcommand{\dDiam}{36}

\newcommand{\mass}{$(1.12 \pm 0.01) \times 10^{19}$}
\newcommand{\dens}{1,280\,$\pm$\,130}
\newcommand{\poro}{50\,$\pm$\,9}

\newcommand{\rem}[1]{\sout{\textcolor{red}{#1}}}

\newcommand{\add}[1]{#1}

\renewcommand{\rem}[1]{}

\journal{Icarus}

\begin{document}

\begin{frontmatter}



\title{Physical, spectral, and dynamical properties of asteroid (107) Camilla and its satellites\tnoteref{obs}}

\tnotetext[obs]{%
  Based on observations obtained at: 1) the Hubble Space Telescope, operated by NASA and ESA; 2) the Gemini Observatory 
  and acquired through the Gemini Observatory Archive, which is
  operated by the Association of Universities for Research in
  Astronomy, Inc., under a cooperative agreement with the NSF on
  behalf of the Gemini partnership: the National Science Foundation
  (United States), the National Research Council (Canada), CONICYT
  (Chile), Ministerio de Ciencia, Tecnolog\'{i}a e Innovaci\'{o}n
  Productiva (Argentina), and Minist\'{e}rio da Ci\^{e}ncia,
  Tecnologia e Inova\c{c}\~{a}o (Brazil); 
  3) the European Southern Observatory, Paranal, Chile --
  \href{http://archive.eso.org/wdb/wdb/eso/sched_rep_arc/query?progid=71.C-0669}{071.C-0669} (PI Merline),
  \href{http://archive.eso.org/wdb/wdb/eso/sched_rep_arc/query?progid=73.C-0062}{073.C-0062} \&
  \href{http://archive.eso.org/wdb/wdb/eso/sched_rep_arc/query?progid=74.C-0052}{074.C-0052}
  (PI Marchis),
  \href{http://archive.eso.org/wdb/wdb/eso/sched_rep_arc/query?progid=87.C-0014}{087.C-0014}
  (PI Marchis),
  \href{http://archive.eso.org/wdb/wdb/eso/sched_rep_arc/query?progid=88.C-0528}{088.C-0528} (PI Rojo),
  \href{http://archive.eso.org/wdb/wdb/eso/sched_rep_arc/query?progid=95.C-0217}{095.C-0217} \&
  \href{http://archive.eso.org/wdb/wdb/eso/sched_rep_arc/query?progid=297.C-5034}{297.C-5034} (PI Marsset)
  --
  and 4) the W. M. Keck Observatory, which is operated as a scientific
  partnership among the California Institute of Technology, the
  University of California and the National Aeronautics and Space
  Administration. The Observatory
  was made possible by the generous financial support of the W. M. Keck
  Foundation.
}


\author[imcce,pucp]{M. Pajuelo}
\author[imcce,oca]{B. Carry}
\author[imcce]{F. Vachier}
\author[qub]{M. Marsset}
\author[imcce]{J. Berthier}
\author[imcce]{P. Descamps}
\author[swri]{W. J. Merline}
\author[swri]{P. M. Tamblyn}
\author[oca,ou]{J. Grice}
\author[lbto]{A. Conrad}
\author[tow]{A. Storrs}
\author[iota-tim]{B. Timerson}
\author[iota-dun]{D. Dunham}
\author[iota-pres]{S. Preston}
\author[lam]{A. Vigan}
\author[eso]{B. Yang}
\author[lam]{P. Vernazza}
\author[bardon]{S. Fauvaud}
\author[cdr]{L. Bernasconi}
\author[cdr]{D. Romeuf}
\author[cdr,geneve]{R. Behrend}
\author[eso,tmt]{C. Dumas}
\author[star]{J. D. Drummond}
\author[ucla]{J.-L. Margot}
\author[uc,lesia]{P. Kervella}
\author[seti]{F. Marchis}
\author[stsci]{J. H. Girard}

\address[imcce]{IMCCE, Observatoire de Paris, PSL Research University, CNRS,
  Sorbonne Universit{\'e}s, UPMC Univ Paris 06, Univ. Lille, France}
\address[pucp]{Secci{\'o}n F{\'i}sica, Departamento de Ciencias, Pontificia Universidad Cat{\'o}lica del Per{\'u}, Apartado 1761, Lima, Per{\'u}}
\address[oca]{Universit\'e C{\^o}te d'Azur, Observatoire de la
  C{\^o}te d'Azur, CNRS, Laboratoire Lagrange, France}
\address[qub]{Astrophysics Research Centre, Queen's University Belfast, Belfast, County Antrim, BT7 1NN, UK}
\address[swri]{Southwest Research Institute, Boulder, Colorado, USA}
\address[ou]{Open University, School of Physical Sciences, The Open University, MK7 6AA, UK}
\address[lbto]{Large Binocular Telescope Observatory, University of Arizona, Tucson, AZ 85721, USA}
\address[tow]{Towson University, Towson, Maryland, USA}
\address[iota-tim]{International Occultation Timing Association (IOTA), 623 Bell Rd., Newark, NY 14513-8805, USA}
\address[iota-dun]{IOTA, 3719 Kara Ct., Greenbelt, MD 20770-3016, USA}
\address[iota-pres]{IOTA, 7640 NE 32 nd St., Medina, WA 98039, USA}
\address[lam]{Aix Marseille Univ, CNRS, LAM, Laboratoire d'Astrophysique de Marseille, Marseille, France}
\address[eso]{ESO-Chile, Alonso de C\'ordova 3107, Vitacura, Santiago, RM, Chile}
\address[bardon]{Observatoire du Bois de Bardon, 16110 Taponnat, France}
\address[cdr]{CdR \& CdL Group: Lightcurves of Minor Planets and Variable Stars, Switzerland}
\address[tmt]{Thirty-Meter-Telescope, 100 West Walnut St, Suite 300, Pasadena, CA 91124, USA}
\address[star]{Leidos, Starfire Optical Range, AFRL/RDS, Kirtland AFB, NM 87117, USA}
\address[geneve]{Geneva Observatory, 1290 Sauverny, Switzerland}
\address[ucla]{Department of Earth, Planetary, and Space Sciences, UCLA, Los Angeles, CA 90095, USA}
\address[uc]{Unidad Mixta Internacional Franco-Chilena de Astronom\'{i}a, CNRS/INSU UMI 3386 and Departamento de Astronom\'{i}a, Universidad de Chile, Casilla 36-D, Santiago, Chile.}
\address[lesia]{LESIA, Observatoire de Paris, PSL Research University, CNRS, Sorbonne Universit\'es, UPMC Univ. Paris 06, Univ. Paris Diderot, Sorbonne Paris Cit\'e, 5 Place Jules Janssen, 92195 Meudon, France.}
\address[seti]{SETI Institute, Carl Sagan Center, 189 Bernado Avenue, Mountain View CA 94043, USA}
\address[stsci]{Space Telescope Science Institute, 3700 San Martin Drive, Baltimore, MD 21218, USA}

\begin{abstract}
  The population of large 100+\,km asteroids is thought to be
  primordial. As such, they are the most direct witnesses of the
  early history of our Solar System available.
  Those among them with satellites allow \rem{to} study \rem{their} \add{of the} mass, \add{and} hence
  density and internal structure.
  We study \add{here} the dynamical, physical, and spectral
  properties of the triple asteroid (107) Camilla from
  lightcurves, stellar occultations, optical spectroscopy, and high-contrast and
  high-angular-resolution images and spectro-images.

  Using \numb{\ObsOne}~positions measured over 15 years, we determine the orbit of
  its \rem{largest} \add{larger} satellite, \iauOne, to be circular,  
  equatorial, and prograde, with root-mean\add{-}square residuals of
  \numb{\rmsOne}~mas, corresponding to a sub-pixel accuracy.
  From \numb{\ObsTwo}~positions spread \rem{in} \add{over} three epochs only\add{,} in 2015 and 2016, we determine a
  preliminary orbit for the second satellite \iauTwo\rem{, slightly tilted and eccentric}\add{.} \add{We 
  find the orbit to be somewhat eccentric and slightly inclined to the
  primary's equatorial plane}, reminiscent of the properties of inner satellites of
  other asteroid triple systems. 
  Comparison of the near-infrared spectrum of \rem{\iauOne} \add{the larger}
  satellite reveals no significant difference with Camilla.
  Hence, both dynamical and surface properties argue for a formation of the
  satellites by excavation from impact and re-accumulation of ejecta in
  orbit. 

  We determine the spin and 3-D
  shape of Camilla. The model fits well each data set: lightcurves,
  adaptive-optics images, and stellar occultations. 
  \add{We determine Camilla to be larger than reported from modeling
    of mid-infrared photometry, with a spherical-volume-equivalent diameter of
  \numb{\Diam\,$\pm$\,\dDiam\,km} (3\,$\sigma$ uncertainty), in agreement
    with recent results from shape modeling
    (Hanus et al., 2017, A\&A 601).}
  Combining the mass \add{of
    \numb{\mass}\,kg (3\,$\sigma$
      uncertainty) determined} from the dynamics of the satellites and the
  volume from the 3-D shape model, we determine a density of
  \numb{\dens}\,kg$\cdot$m$^{-3}$ (3 $\sigma$ uncertainty).
  \add{From this density, and considering Camilla's spectral
    similarities with (24) Themis and (65) Cybele (for which water ice
    coating on surface 
    grains was reported), we infer a silicate-to-ice mass ratio of
    \numb{1--6}, with a \numb{10-30}\% macroporosity.}
\end{abstract}


\end{frontmatter}



\section{Introduction}
  \indent \rem{The asteroids orbiting in the main belt, between Mars and
  Jupiter,}
  \add{Main belt asteroids}
  are the remnants of the building blocks that accreted to form
  terrestrial planets, leftovers of the dynamical events that
  shaped our planetary system.
  Among them, large bodies (diameter larger than $\approx$100\,km) are
  deemed primordial \citep{2009-Icarus-204-Morbidelli}, and contain a
  relatively pristine record of their initial formation conditions. 

  \indent Decades of photometric and spectroscopic surveys
  have provided \rem{a clear view} \add{an ever-improving picture} of the
  distribution of material in the inner solar system  
  \citep[e.g.][]{1982-Science-216-Gradie, 
    1996-MPS-31-Burbine, 2002-AsteroidsIII-5.2-Burbine, 
    2002-Icarus-158-BusII, 
    2002-AsteroidsIII-2.2-Rivkin, 2006-Icarus-185-Rivkin, 
    2008-Nature-454-Vernazza, 2010-Icarus-207-Vernazza,
    2014-ApJ-791-Vernazza,  
    2014-Nature-505-DeMeo}, yet
  these studies have probed the composition of the surface only. 
  As such, they
  \rem{failed to address}
  \add{do not necessarily lead us to}
  the original location and time
  scales for the accretion of these blocks, which are key to
  understand\add{ing} the \add{important} processes \rem{that occurred} in the disk of gas and dust
  around the young Sun. 

  \indent \rem{Fortunately, t} \add{T}hese \rem{questions} \add{issues} can be addressed by studying
  the internal structure of asteroids:
  objects formed far from the Sun are expected to be composed \rem{by a} 
  \add{of various}
  mixture\add{s} of \rem{rocks and ices} \add{rock and ice}, while
  \rem{innermost} objects \add{closer to the Sun} are \rem{deemed}
  \add{expected to be} volatile-free.  
  Depending on their formation time scale, the amount of radiogenic
  heat \rem{ was different} \add{varied}, leading to \add{complete,}
  partial\add{,} or \rem{complete} \add{no} differentiation\rem{,
  or not at all}. 
  In that respect, density is \rem{maybe} \add{clearly} the \rem{main}
  \add{most important} \rem{physical property} remotely
  measurable \add{property} that \add{can} constrain\rem{s} internal structure
  \citep{2015-AsteroidsIV-Scheeres}.  
  
  \indent Determination of \rem{the} density \rem{relies on the}
  \add{requires} measurement of \rem{the}
  mass and \rem{the} volume, and for that, large asteroids with satellites
  are prime targets \citep{1999-Nature-401-Merline,
    2002-AsteroidsIII-2.2-Merline, 
    2008-Icarus-195-Marchis, 2008-Icarus-196-Marchis, 
    2011-AA-534-Carry, 2015-AsteroidsIV-Margot}.
  The study of \rem{their mutual orbit} \add{the orbits of satellites
    within asteroid binaries or
  multiple systems} is currently the most precise method
  to estimate \rem{asteroid masses,} \add{the mass of the primary
    asteroid.} \rem{while they usually} \add{If the primary also
    happen to} have \add{an} angular
  \rem{diameters} \add{diameter} large enough to be spatially resolved by large telescopes,
  \rem{allowing the} \add{this also allows an} accurate determination
  of \rem{their} \add{the primary's} volume. In addition, the orbits of the
  satellites themselves offer a way to probe the gravity field,
  related to mass distribution inside the asteroid
  \citep{2014-Icarus-239-Berthier, 2014-ApJ-783-Marchis}. 

  \indent \rem{We focus in the present study}
  \add{Here we focus} on the outer\add{-}main-belt asteroid (107)
  Camilla,
  \add{orbiting in the Cybele region} \rem{asteroid,} \add{and} discovered on November 17, 1868 from Madras, 
  India by N. R. Pogson. Its first satellite, \iauOne~\add{(hereafter
  \SatOne)}, was
  discovered in March 2001 by
  \citet{2001-IAUC-Storrs}, using the Hubble Space Telescope (HST),
  and its orbit first studied by \citet{2008-Icarus-196-Marchis} using
  observations from large ground-based telescopes equipped with
  adaptive-optics (AO) systems.
  Its second satellite, \iauTwo~\add{(hereafter
  \SatTwo)}, was discovered in 2016 by our team
  \citep{2016-IAUC-Marsset}, using the European Southern Observatory
  (ESO) Very Large Telescope (VLT). 

  \indent \add{
    Camilla was originally classified as a C-type
    based on its visible colors and albedo
    \citep{1989-AsteroidsII-Tedesco}.
    Later on, both \citet{2002-Icarus-158-BusII} and
    \citet{2004-Icarus-172-Lazzaro} classified it as 
    X, based on visible spectra. More recently, based on a
    near-infrared spectrum from NASA IRTF Spex, 
    \citet{2015-Icarus-247-Lindsay} classified Camilla as either Xe or
    L. }

  \indent \add{
    The physical properties of Camilla have been extensively studied, 
    from its rotation period of \numb{4.8}\,h
    \citep[e.g.,][]{1987-Icarus-70-Weidenschilling,
      1987-Icarus-69-DiMartino} to its spin and 3D shape model
    \citep[][]{2003-Icarus-164-Torppa, 
      2011-Icarus-214-Durech, 
      2013-Icarus-226-Hanus, 
      2017-AA-601-Hanus}.
    Its diameter, however, was poorly constrained, with estimates
    ranging from
    185\,$\pm$\,9\,km \citep{2006-Icarus-185-Marchis} to 
    256\,$\pm$\,12\,km \citep{2012-Icarus-221-Marchis}. More 
    recent studies combining 
    images or stellar occultations with lightcurve-based 3D shape
    modeling, are yielding 
    diameters in excess of 220\,km
    (see Fig.~\ref{fig:diam} and Table~\ref{tab:diam} for the exhaustive
    list of diameter estimates).
    The mass estimates 
    also spanned a wide range, from
    \numb{$2.25_{-2.25}^{+18.00}$} to
    \numb{$39 \pm 10 \times 10^{18}$\,kg} \citep{2011-AJ-142-Zielenbach} 
    (see Fig.~\ref{fig:mass} and Table~\ref{tab:mass} for the
    exhaustive list of mass estimates).
    With these large spread of values, deriving an
    accurate density would require substantial
    improvements to these parameters.
  }
  
  \indent Gathering all the available disk-resolved and high-contrast images
  from HST and AO-fed cameras, optical lightcurves, stellar occultations,
  and visible and near-infrared spectra
  (Section~\ref{sec:obs}), we present an extensive study of the dynamics of  
  the system (Section~\ref{sec:dyn}), of the surface properties of Camilla and
  its main satellite \add{\SatOne}~(Section~\ref{sec:spec}), and of Camilla's spin and 3-D
  shape (Section~\ref{sec:phys}), all constraining its internal
  \add{composition and} structure (Section~\ref{sec:discuss}).

\section{Observations\label{sec:obs}}

  \subsection{Optical lightcurves\label{ssc:obs:lc}}

    \indent We gather the 24 lightcurves used by
    \citet{2003-Icarus-164-Torppa} to create a convex 3-D shape model of
    Camilla\footnote{Available on DAMIT \citep{2010-AA-513-Durech}:\\
      \href{http://astro.troja.mff.cuni.cz/projects/asteroids3D/}{http://astro.troja.mff.cuni.cz/projects/asteroids3D/}},
    compiled from the Uppsala Asteroid Photometric
    Catalog\footnote{\href{http://asteroid.astro.helsinki.fi/apc/asteroids/}{http://asteroid.astro.helsinki.fi/apc/asteroids/}}
    \citep{PDS-SAPC-2011}. We also retrieve the three lightcurves reported by
    \citet{2009-MPBu-Polishook}. 

    \indent In addition to these data, we acquired \numb{29}
    lightcurves using the  
    60\,cm \textsl{Andr{\'e} Peyrot} telescope mounted at Les Makes
    observatory on R{\'e}union Island, \rem{which is} operated as a 
    partnership among Les Makes Observatory and the IMCCE, Paris
    Observatory. 
    We also extracted \numb{63} lightcurves from the data archive of
    the SuperWASP survey \citep{2006-PASP-118-Pollacco} for the
    period 2006-2009.
    This survey aims \rem{at finding and characterizing} \add{to find
      and characterize} exoplanets by
    observation\add{s} of their transit\add{s of} \rem{in front of
      their} \add{the} host star.
    Its large field of view (8\degr\,$\times$\,8\degr) provides a 
    goldmine for asteroid lightcurves \citep{2005-EMP-97-Parley,
      2017-ACM-Grice}.

    \indent A total of \numb{\nbLC} lightcurves observed between
    \numb{1981} and \numb{2016}
    (Table~\ref{tab:lc}) are used in this work.

  \subsection{High-angular-resolution imaging\label{ssc:obs:ao}}

    \indent We compile here all the high-angular-resolution images of \rem{(107)}
    Camilla taken with the HST and 
    large ground-based telescopes equipped with AO-fed
    cameras: Gemini North, ESO VLT, and W. M. Keck, of which only a subset had already been published
    \citep{2001-IAUC-Storrs, 2008-Icarus-196-Marchis}.
    All of these data sets were acquired by the authors
    of this paper.  The data comprise \numb{62} different epochs, with
    multiple images each, spanning \numb{15} years, from March 2001 to
    August 2016. 

    \indent The images from the VLT were acquired with
    both the first generation instrument NACO
    \citep[NAOS-CONICA,][]{2003-SPIE-4841-Lenzen,2003-SPIE-4839-Rousset} and  
    SPHERE
    \add{\citep[Spectro-Polarimetric High-contrast Exoplanet REsearch,][]{2006-OExpr-14-Fusco,2008-SPIE-Beuzit}}, the
    second generation extreme-AO instrument designed for 
    exoplanet detection and characterization.
    The images taken with SPHERE used its IRDIS differential imaging
    camera sub-system
    \add{\citep[InfraRed Dual-band Imager and Spectrograph,][]{2008-SPIE-Dohlen}}.
    Images taken at the Gemini North used NIRI camera
    \add{\citep[Near InfraRed Imager,][]{2003-PASP-115-Hodapp}}, fed by the ALTAIR AO system
    \citep{2000-AOST-Herriot}. Finally, observations at Keck were
    \rem{realized} \add{acquired} 
    with \add{NIRC2
    \citep[Near-InfraRed Camera 2,][]{2004-AppOpt-43-vanDam,2000-SPIE-4007-Wizinowich}}.
    We list in Table~\ref{tab:ao} the details of each observation. 

    \indent The basic data processing (sky subtraction, bad-pixel
    removal, and flat-field 
    correction) was performed using
    in-house routines developed in Interactive Data Language (IDL) to
    reduce AO-imaging data 
    \citep[see][for more details]{2008-AA-478-Carry}.

  \subsection{High-angular-resolution spectro-imaging\label{ssc:obs:ifu}}

    In 2015 and 2016, we also used the \add{integral-field
      spectrograph (IFS)} of the SPHERE instrument
    at the ESO VLT, aiming \rem{at measuring} \add{to measure} the
    reflectance spectrum of Camilla's largest satellite \SatOne, and the
    astrometry of the fainter satellite \SatTwo.
    The observations were \rem{carried out} \add{made} in the IRDIFS\_EXT mode
    \citep{2014-AA-572-Zurlo}, in
    which both IRDIS \citep{2008-SPIE-Dohlen} and the IFS \citep{2008-SPIE-Claudi}
    \add{data are acquired} 
    \rem{observe} simultaneously. In
    this set-up, the IFS covers the wavelength range from 0.95 to
    1.65\,$\mu$m (YJH bands) at a spectral resolving power of $\sim$30 in
    a 1.7\arcsec $\times$1.7\arcsec~field of view (FoV),
    while IRDIS \rem{observes} \add{operates} in the dual-band imaging mode
    \citep[DBI,][]{2010-MNRAS-407-Vigan}
    with $K_{12}$, a pair of filters in the {\it K} band ($\lambda_{K_1}$ =
    2.110\,$\mu$m and $\lambda_{K_2}$ = 2.251\,$\mu$m, $\sim$0.1\,$\mu$m
    bandwidth), within a 4.5\arcsec~FoV.
    All observations were performed in the pupil-tracking mode,
    where the pupil remains fixed while the field orientation varies during the
    observations. This mode provides the best PSF stability and helps in reducing
    and subtracting static speckle noise in the images. 

    \indent For the pre-processing of both the IFS and IRDIS data,
    we \rem{made use of} \add{used} the preliminary release (v0.14.0-2) of the SPHERE
    Data Reduction and Handling (DRH) software
    \citep{2008-SPIE-7019-Pavlov}, as well as additional in-house tools written
    in IDL, including parts of the public pipeline presented in
    \citet{2015-MNRAS-454-Vigan}.
    {See our recent works on (3) Juno and
    (6) Hebe for more details
    \citep{2015-AA-581-Viikinkoski, 2017-AA-604-Marsset}.}
    We used the DRH for the creation of some of the basic
    calibrations: master sky frames, master flat-field, IRDIFS spectra positions,
    initial wavelength calibration and flat field. Before creating the data
    cubes, we used IDL routines to subtract the background from each science frame
    and correct for the bad pixels identified using the master dark and master
    flat-field DRH products. This step was introduced as a substitute
    \rem{to} \add{for} the bad
    pixel correction provided by the DRH. Bad pixels were first identified using a
    sigma-clipping routine, and then corrected using a bicubic pixel
    interpolation with the \texttt{MASKINTERP} IDL routine. The resulting frames were then
    injected into the DRH recipe to create the data cubes by interpolating the
    data spectrally and spatially.

  \subsection{Stellar occultations\label{sec:obs:occ}}

    \indent Eleven stellar occultations by Camilla have been observed
    in the last decade, mostly by amateur astronomers
    \citep[see][]{2014-ExA-38-Mousis, 2016-IAU-Dunham}.
    The timings of disappearance and reappearance of the
    stars, together with the \rem{localization} \add{location} of each observing station
    are compiled by \citet{PDSSBN-OCC}, and publicly available on the Planetary Data
    System
    (PDS\footnote{\href{http://sbn.psi.edu/pds/resource/occ.html}{http://sbn.psi.edu/pds/resource/occ.html}}).
    We converted the disappearance and reappearance timings
    (Table~\ref{tab:obsocc})
    of the occulted stars into segments (called chords) on the plane
    of the sky, using the location of the observers on Earth and the
    apparent motion of Camilla following the recipes by \citet{1999-IMCCE-Berthier}.
    Four stellar occultations had multiple chords\add{;}\rem{, 
    while the} other 
    events \add{had} only \rem{had} one or two positive chords, and
    \rem{provided}  
    \add{contributed less to constraining} \rem{constraints on} the size and apparent shape of Camilla.
    \add{In n}\rem{N}one of these eleven stellar occultations was
    \add{there any evidence for} \rem{indicative of the
    presence of} a companion.
    We list in Table~\ref{tab:occ} the details of the seven events that
    we used. 

%
%

  \subsection{Near-infrared spectroscopy\label{sec:obs:spex}}

    \indent On November 1, 2010, we observed Camilla over
    0.8--2.5\,$\mu$m with the  
    near-infrared spectrograph SpeX \citep{2003-PASP-115-Rayner}, on
    the 
    3-meter NASA IRTF located on Mauna Kea, Hawaii, 
    using the low resolution Prism mode ($R$\,=\,100). We used the
    standard \textsl{nodding} procedure for the observations, using
    alternately two separated locations on the slit
    \citep[e.g.,][]{2007-AA-470-Nedelcu} to estimate the sky background.
    We used Spextool (SPectral EXtraction TOOL), an
    IDL-based data reduction package written by
    \cite{2004-PASP-116-Cushing} to reduce SpeX data.

\begin{figure*}[ht]
  \includegraphics[width=\textwidth]{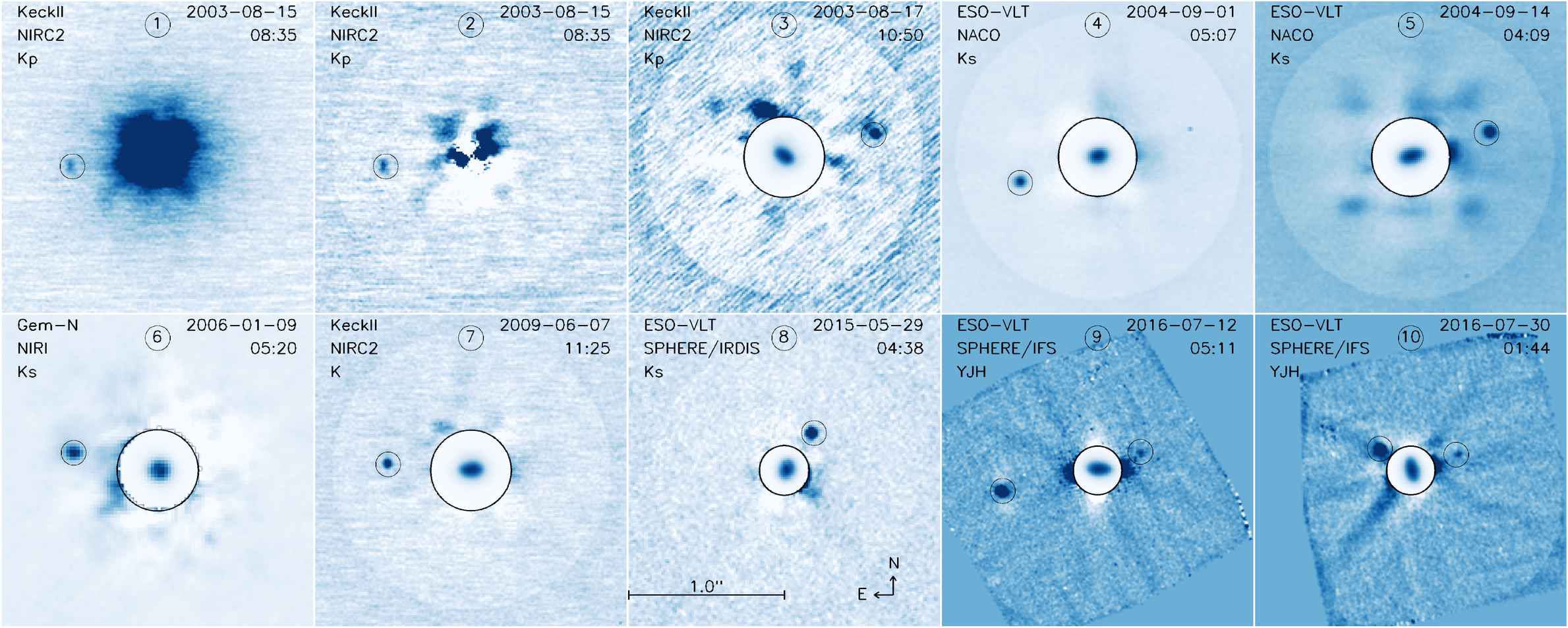}
  \caption[Examples of AO images of Camilla]{
    Examples of AO images from Gemini, Keck, and ESO VLT.
    The first two panels (\add{1 \& 2}, August 13, 2003, from Keck) show a typical
    AO image, before and after halo subtraction: 
    Camilla dominates the background and makes the satellites hard to
    detect.
    The remaining panels show halo-subtracted images from different
    dates, with small circles indicating the positions of
    \add{the bright satellite \SatOne~and the fainter
      \SatTwo~(frames 9 and 10 only)}.
    On these panels, the images before subtraction are also
    shown in the central circle to highlight the elongated shape of Camilla.
  }
  \label{fig:ao}
\end{figure*}

\section{Dynamical properties\label{sec:dyn}}

  \subsection{Data processing\label{sec:dyn:data}}

    \indent The main challenges in measuring the position and
    apparent flux of the satellite of an
    asteroid results from their sub-arcsecond angular separation and
    high contrast (several magnitudes), combined with
    \rem{non-perfect} \add{imperfect} AO correction.
    A typical image of a binary asteroid (Fig.~\ref{fig:ao}) displays
    a central peak (the asteroid itself, angularly resolved or not)
    encompassed by a halo (its diffused light), 
    within which speckle patterns appear.
    The faintness of these speckles, produced by 
    interference of the incoming light, make them very similar in
    appearance to a small moon with a contrast up to several thousands,
    and they can be misleading. 
    Speckles, however, \add{vary (position and flux)} on
    short timescales, depending on the ambient conditions
    and AO performances (e.g., seeing, airmass, brightness of the AO
    reference source).
    These fluctuations can be used to distinguish genuine satellites from
    speckles. 

    \indent As for the direct imaging of exoplanets, it is crucial to
    substract the halo that surrounds the primary
    \citep[in a similar way to the digital coronography of][]{2008-PSS-56-Assafin}.
    Because asteroids
    are also marginally resolved, their light is not fully coherent,
    and the speckle pattern is not as stable in time, nor simple, as
    in the case of a star. 
    The tool we developed considers concentric annuli around the center
    of light of the primary to evaluate its halo. Although the principle is
    straightforward, great caution was taken in the implementation, especially
    in the computation of the
    intersection of the annulus with the pixels to allow the use of
    annuli with a sub-pixel width. The contribution of each pixel to
    different annuli is thus solved first, and the median flux of each
    annulus is computed, and subtracted from each pixel accordingly.

    \indent The position and flux of the satellite, relative to the
    primary, is then measured by fitting a 2-D Gaussian function to
    the halo-subtracted image. 
    \add{The satellites are distinguished from
    speckles by comparing different images, taken both
    close in time and over a range of times.} 
    To estimate the uncertainties on the
    position and apparent flux of both the primary and the satellites,
    we use different \add{integration} apertures for \add{each object.} \rem{which their size is determined by
    the fit the} \add{The sizes of the apertures are
    determined by fitting a} 2-D Gaussian \rem{function} \add{to
      each}, \add{with diameters} typically \rem{from} \add{being} 
    5 to 150 pixels for the primary, and 3 to 15 \add{pixels} for the satellites.
    The reported positions and apparent magnitudes
    (Tables~\ref{tab:genoid1} and \ref{tab:genoid2})
    are the average of all fits (after removal of outlier values), and the
    reported uncertainties are the standard deviations.

  \subsection{Orbit determination with \genoid\label{sec:dyn:genoid}}

    \indent We use our algorithm \genoid~\citep[GENetic Orbit
      IDentification,][]{2012-AA-543-Vachier} 
    to determine the orbit of the satellites. \genoid~is a genetic-based
    algorithm that relies on a metaheuristic method to find the
    best-fit (i.e., minimum $\chi^2$) suite of dynamical parameters
    (\add{mass, semi-major axis, eccentricity, inclination, longitude
      of the node, argument of pericenter, and time of passage to
      pericenter}) by 
    refining, generation after generation, a grid of test values
    (called \textsl{individuals}). 

    \indent The first generation is drawn randomly over a very
    wide range for each parameter, \add{thus avoiding a miss of the
    global minimum from inadequate initial conditions}\rem{, which is always
    a threat in minimization algorithms}.
    For each individual (i.e., set of dynamical
    parameters), the $\chi^2$ residuals between the observed and predicted
    positions is computed as
    \begin{equation}
      \label{eq:chi2}
      \chi^2 = \sum_{i=1}^{N} \left[
        \left(\frac{X_{o,i} - X_{c,i}}{\sigma_{x,i}} \right)^2 + 
        \left(\frac{Y_{o,i} - Y_{c,i}}{\sigma_{y,i}} \right)^2 \right]
    \end{equation}
    
    \noindent where $N$ is the number of observations, and $X_i$ and $Y_i$
    are the relative positions between the satellite and Camilla along
    the right ascension and declination respectively. 
    The indices $o$ and $c$ stand for observed and computed
    positions, and $\sigma$ are the measurement uncertainties.

    \indent A new generation of individuals is drawn
    by mixing randomly the parameters of individuals with the lowest
    $\chi^2$ from the former generation\rem{, in a survival of the fittest
    fashion}. 
    This way, the entire parameter space is scanned, with the
    density of evaluation points increasing toward low $\chi^2$
    regions along the process.
    At each generation, we also use the best
    individual as initial condition to search for the
    local minimum by gradient descent.
    The combination of genetic grid-search and gradient descent thus
    ensures finding \textsl{the} best solution. 

    \indent We then assess the confidence interval of the dynamical
    parameters by considering all the individuals providing predictions
    within 1, 2, and 3\,$\sigma$ of the observations. The range
    spanned by these individuals provide the
    confidence interval at the corresponding $\sigma$ level for each
    parameter. 

    \indent The reliability of \genoid~has been assessed during a stellar
    occultation by (87) Sylvia and its satellites Romulus and Remus on
    January 6, 2013:  
    \add{\genoid~had been used to predict the position of Romulus
      before the event, directing observers to locations specifically to
      target the satellite. Four different observers detected
      an occultation by Romulus at only 13.5\,km
      off the predicted track  
      \citep[the cross-track uncertainty was 65\,km,][]{2014-Icarus-239-Berthier}. }

  \subsection{Orbit of \SatOne: \iauOne\label{sec:dyn:S1}}

    \indent We measured \numb{\ObsOne} astrometric positions of the
    satellite \SatOne~relative to Camilla over a span of \numb{15} years,
    corresponding to \numb{5642} days or \numb{1520} revolutions.
    The orbit we derive with \genoid~fits all \numb{\ObsOne} observed
    positions of the satellite with a root mean square (RMS) residual
    of \numb{\rmsOne} milli-arcseconds (mas) only, which corresponds to
    a sub-pixel accuracy.

\begin{figure}[ht]
  \includegraphics[width=.5\textwidth]{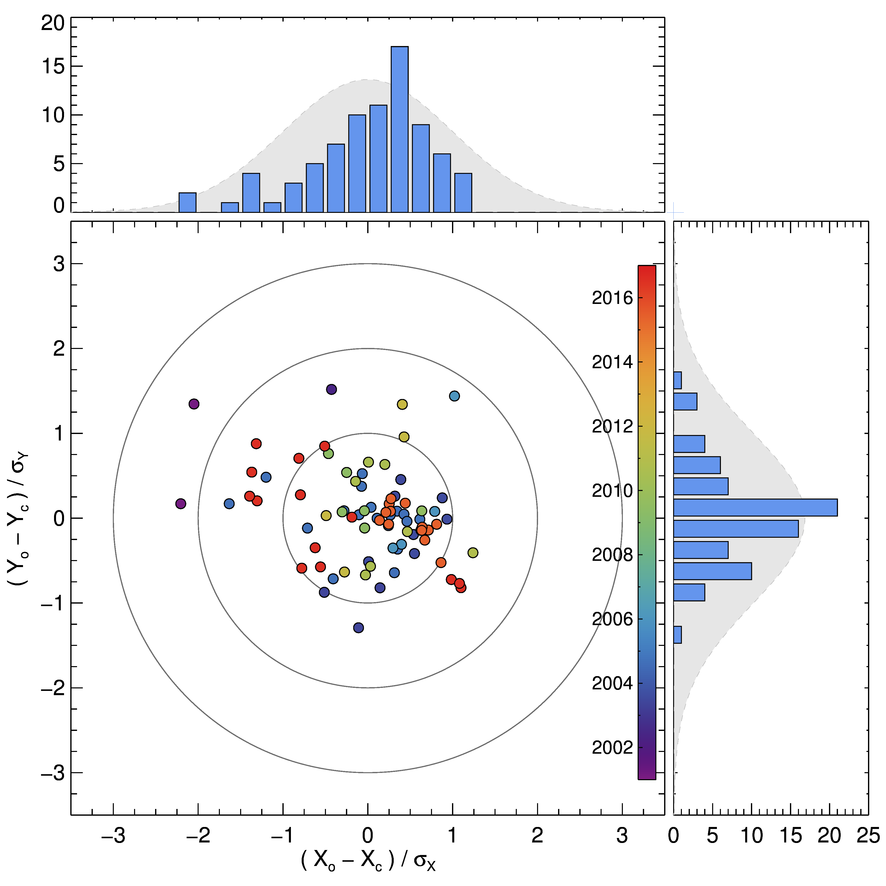}
  \caption[Residuals on predicted positions of \SatOne]{
    Distribution of residuals \add{for \SatOne} between the
    observed (index o)
    and
    predicted (index c) positions, normalized by the uncertainty on
    the measured 
    positions ($\sigma$), and color-coded by observing epoch.
    X stands for right ascension and Y for declination.
    The three large gray circles represent the 1, 2, and 3 $\sigma$
    limits. 
    The top panel shows the histogram of residuals along X, and the
    right panel the residuals along Y. The light gray Gaussian in the
    background has a standard deviation of \rem{1} \add{one}.
  }
  \label{fig:bin:rms}
\end{figure}

    \indent \SatOne~orbits Camilla on a circular, prograde, equatorial
    orbit, in \numb{\pOne} days with a semi-major axis of \numb{\aOne}\,km.
    We detail all the parameters of its orbit in
    Table~\ref{tab:dyn}, with their confidence interval taken at
    3\,$\sigma$. The distribution of residuals between the
    observed and predicted positions, normalized by the uncertainty on
    the measured positions, are plotted in Fig.~\ref{fig:bin:rms}. 
    The orbit we determine here is qualitatively similar to the one
    \add{given} by 
    \citet{2008-Icarus-196-Marchis}, while much better constrained: 
    we fit \numb{\ObsOne} astrometric positions over \numb{15} years
    with \rem{a} \add{an} RMS residual
    of \numb{\rmsOne} mas, compared to their fit of
    \numb{23} positions over less than \numb{3} years with \rem{a} \add{an} RMS
    residual of \numb{22} mas. 
    \add{The much longer time span of observations provides a much 
      more stringent 
      constraint on the period
      (\numb{3.712\,34\,$\pm$\,0.000\,04\,day}) of \SatOne, 
      compared to the value of 
      \numb{3.722\,$\pm$\,0.009\,day} reported by
      \citet{2008-Icarus-196-Marchis}.}

    \indent \add{As a result, we determine a much more precise mass for Camilla of
      \numb{$(1.12 \pm 0.01) \times 10^{19}$}\,kg (3\,$\sigma$
      uncertainty), about 1\% of the mass of Ceres
      \citep{2012-PSS-73-Carry}.
      We list in Table~\ref{tab:mass} the reported values of the mass
      of Camilla found in the literature. Our mass value agrees well
      with the average value 
      \numb{$(1.10 \pm 0.69) \times 10^{19}$}\,kg we show in Table~\ref{tab:mass}, although
      the mass estimates derived from orbital deflection and
      solar system ephemerides have a large scatter 
      \citep[see][for a discussion on the precision and bias of mass
        determination methods]{2012-PSS-73-Carry}.
      Our determination significantly reduces the
      uncertainty in the prior 
      value of \numb{$(1.12 \pm 0.09) \times 10^{19}$}\,kg, that also
      used the orbit of \SatOne~\citep{2008-Icarus-196-Marchis}.
    }

\begin{figure}[ht]
  \includegraphics[width=.5\textwidth]{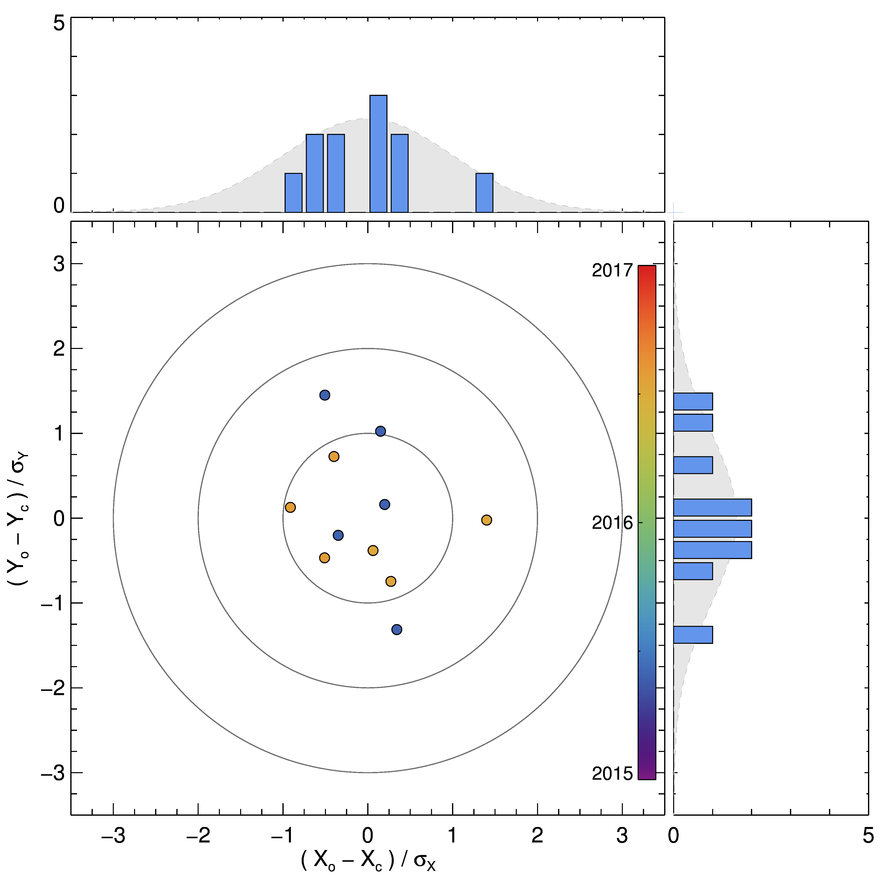}
  \caption[Residuals on predicted positions of \SatTwo]{
    Similar \rem{as} \add{to} Fig.~\ref{fig:bin:rms}, but for \SatTwo.
  }
  \label{fig:bin:rms2}
\end{figure}

%
%
%

%
%
  \subsection{Orbit of \SatTwo: \iauTwo\label{sec:dyn:S2}}

    \indent We measured \numb{\ObsTwo} astrometric positions of the
    satellite \SatTwo~relative to Camilla \rem{between} \add{during} \numb{2015
    and 2016}, corresponding to \numb{428} days or \numb{311} revolutions.
    \rem{Unfortunately,} \rem{these} \add{These} observations correspond to \rem{only} three
    well-separated epochs: 2015-May-29, 2016-Jul-12, and 2016-Jul-30,
    providing \add{the minimum needed to constrain} \rem{little constraints on} the orbit.
    \add{Thus, although the orbit we determine with \genoid~fits all
      \numb{\ObsTwo} observed positions of \SatTwo~with an RMS
      residual of only \numb{\rmsTwo} mas and yields
      reliable values for the major orbital elements,
      details of all orbital parameters will require further
      observations.} 

\begin{figure}[ht]
  \includegraphics[width=.5\textwidth]{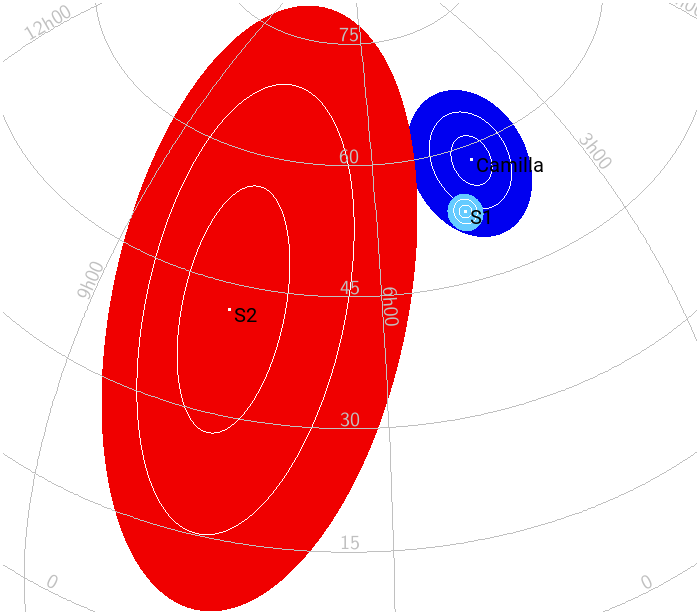}
  \caption[Spin location of the orbits and Camilla]{
    Coordinates and 1 -- 2 -- 3 $\sigma$ contours of Camilla\add{'s} spin axis
    (blue) and the orbital poles of \SatOne~(gray) and \SatTwo~(red)
    in ecliptic coordinates.
  }
  \label{fig:spins}
\end{figure}

    \indent \add{
      \SatTwo~orbits Camilla in
      \numb{\pTwo} days with a semi-major axis of \numb{\aTwo}\,km.
      We detail all the parameters of its orbit in
      Table~\ref{tab:dyn} and present the distribution of residuals
      between the observed and predicted positions in Fig.~\ref{fig:bin:rms2}.
      Unlike \SatOne, its orbit seems neither equatorial nor
      circular.
      While cognizant of the larger uncertainties,
      we favor an orbit inclined to the equator
      of Camilla by an angle 
      $\Lambda$ of \numb{32\,$\pm$\,28}\degr~(Fig.~\ref{fig:spins}),
      and a more eccentric orbit (\numb{e=\eTwo$_{-0.18}^{+0.23}$}).  
      Although a circular orbit, co-planar with \SatOne~is marginally
      within the range of uncertainty, 
      such a solution results in significantly higher residuals.
      This configuration of an outer satellite on a circular and
      equatorial orbit with an inner satellite 
      on an inclined and more eccentric orbit 
      has already been reported for other
      triple systems:
      (45) Eugenia, (87) Sylvia, and (130) Elektra
      \citep{2010-Icarus-210-Marchis, 2012-AJ-144-Fang,
        2014-Icarus-239-Berthier, 2016-ApJ-820-Yang,
        2016-Icarus-276-Drummond}. 
    }

\begin{table*}
\begin{center}
  \caption[Orbital elements of the satellites of Camilla]{Orbital elements of the satellites of Camilla\add{,}
    \SatOne~and \SatTwo, expressed in EQJ2000,
    obtained with \genoid:
    orbital period $P$, semi-major axis $a$,
    eccentricity $e$, inclination $i$,
    longitude of the ascending node $\Omega$,
    argument of pericenter $\omega$, time of pericenter $t_p$.
    The number of observations and RMS between predicted and
    observed positions are also provided.
    \rem{We finally} \add{Finally, we} report the derived primary mass $M$,
    the ecliptic J2000 coordinates of the orbital pole
    ($\lambda_p,\,\beta_p$), 
    the equatorial J2000 coordinates of the orbital pole
    ($\alpha_p,\,\delta_p$), and the
    \rem{angular tilt} \add{orbital inclination} ($\Lambda$) with respect to the equator of
    Camilla. Uncertainties are given at 3-$\sigma$.}
  \label{tab:dyn} 
  \begin{tabular}{l ll ll}
    \hline\hline
     & \multicolumn{2}{c}{\SatOne} & \multicolumn{2}{c}{\SatTwo} \\
    \hline
    \noalign{\smallskip}
    \multicolumn{2}{c}{Observing data set} \\
    \noalign{\smallskip}
    Number of observations & \ObsOne   & & \ObsTwo \\
    Time span (days)       & 5642 &      & 428 \\
    RMS (mas)              & \rmsOne   &      & \rmsTwo \\
    \hline
    \noalign{\smallskip}
    \multicolumn{2}{c}{Orbital elements EQJ2000} \\
    \noalign{\smallskip}
    $P$ (day)        &      3.712\,34  & $\pm$ 0.000\,04  & 1.376  &  $\pm$ 0.016 \\
    $a$ (km)         &     1247.8    & $\pm$ 3.8      &  643.8&  $\pm$ 3.9 \\
    $e$              &       0.0     & $+$  0.013     & 0.18  & $_{-0.18}^{+0.23}$  \\
    $i$ (\degr)      &      16.0     & $\pm$  2.3     & 27.7  & $\pm$ 21.8 \\
    $\Omega$ (\degr) &      140.1    & $\pm$  4.9     & 219.9 & $\pm$ 67.0 \\
    $\omega$ (\degr) &       98.7    & $\pm$  6.5     & 199.4 & $\pm$ 37.6 \\
    $t_{p}$ (JD)     & 2452835.902   & $\pm$ 0.067     & 2452835.31589 & $\pm$ 0.174   \\
    \hline
    \noalign{\smallskip}
    \multicolumn{2}{c}{Derived parameters} \\
    \noalign{\smallskip}
    $M$ ($\times 10^{19}$ kg)     &  1.12   & $\pm$ 0.01    \\
    $\lambda_p,\,\beta_p$ (\degr) & 73, +53   & $\pm$ 2, 2  & 114, +42 & $\pm$ 44, 18   \\
    $\alpha_p,\,\delta_p$ (\degr) & 50, +74   & $\pm$ 5, 2  & 130, +62 & $\pm$ 67, 22   \\
    $\Lambda$ (\degr)             &  4        & $\pm$ 8    &  32 &   $\pm$ 28 \\
    \hline
  \end{tabular}
\end{center}
\end{table*}


\section{Surface properties\label{sec:spec}}

  \subsection{Data processing\label{sec:spec:data}}

    \indent We measured the near-infrared spectra of Camilla and its largest
    satellite \SatOne~using the SPHERE/IRDIFS data.
    Telluric features were removed, and the reflectance spectra were
    obtained by observing the nearby solar type star HD139380.

    \indent Similarly to previous sections, the bright halo of Camilla that 
    contaminated the spectrum of the moon was removed.
    This was achieved by measuring the
    background at the location of the moon for each pixel 
    as the median value of the area defined as a
    40$\times$1-pixel arc \rem{centred} \add{centered} on Camilla.
    To estimate the uncertainty and potential bias on photometry
    introduced by this method, we performed a 
    number of simulations in which we injected fake companions on the 39 spectral
    images of the spectro-imaging cube, at separation ($\approx$300
    mas) and random position angles from the
    primary. The simulated sources were modeled as the PSF, from the
    calibration star images, scaled in brightness. 

\begin{figure}[ht]
  \includegraphics[width=.5\textwidth]{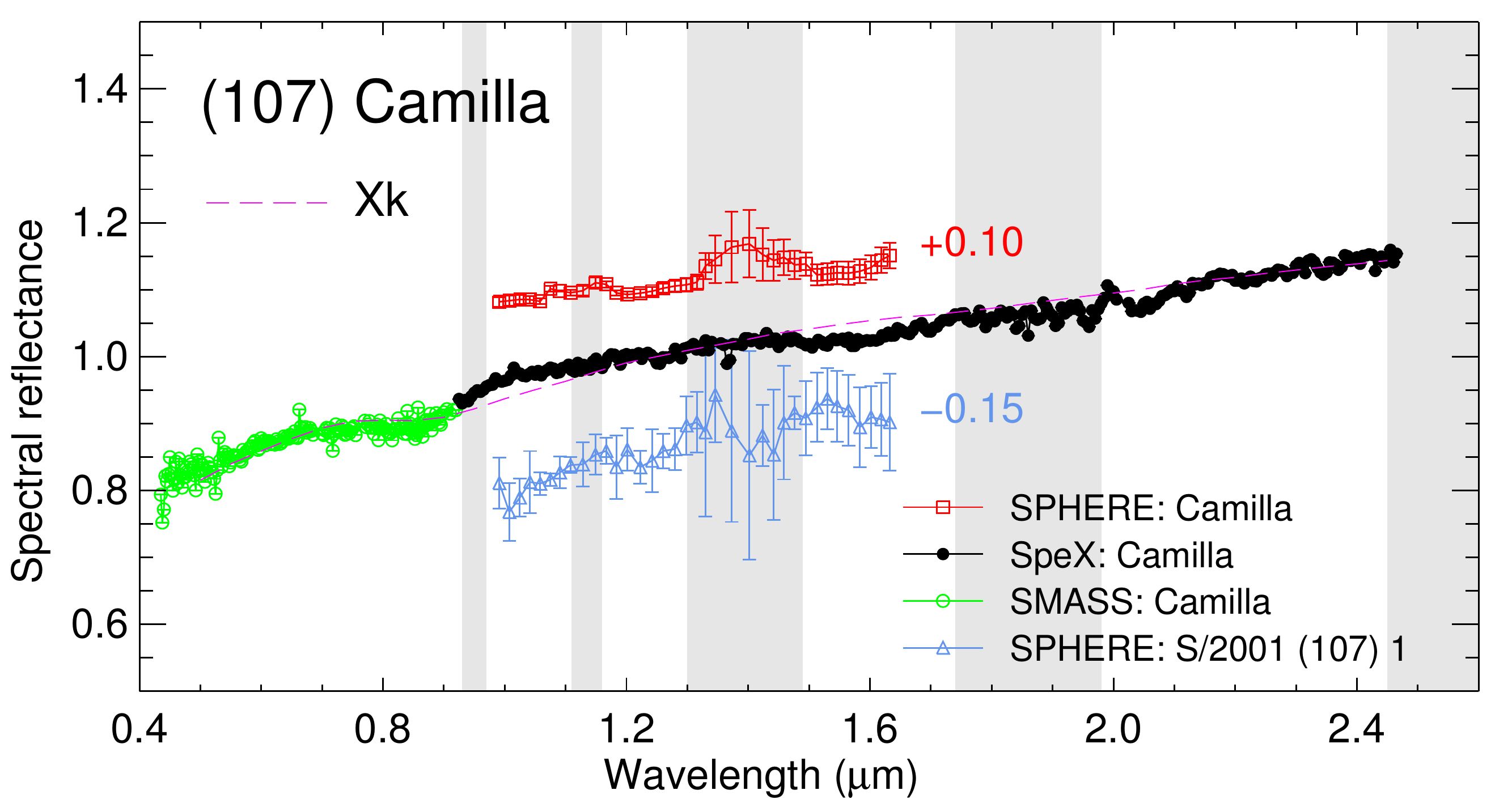}
  \caption[Near-infrared spectra of Camilla and its largest moon \SatOne]{
    Visible and near-infrared spectrum of Camilla from IRTF (green
    and black dots) and
    SPHERE (red squares, offset by +0.1),
    and its moon \SatOne~from SPHERE (blue triangles, offset by -0.15).
    Gray areas represent the wavelength ranges affected by water
    vapour in the atmosphere.
    All spectra were normalize\add{d} to unity at one micron.
    Overplot to the IRTF spectra is the Bus-DeMeo Xk class.
  }
  \label{fig:spec}
\end{figure}

    \indent The halo from Camilla was then removed from these
    simulated images using the method described above, and the flux of the simulated
    companion measured by adjusting a 2D-Gaussian profile.
    Based on a total statistics of 500 simulated companions, we find
    that the median loss of flux at each wavelength is 
    \numb{11$\pm$10\%}. A spectral gradient is also introduced by our
    technique, but it is smaller than \numb{0.06$\pm$0.07\%\,$\cdot$$\mu$m$^{-1}$}.
    The spectra of Camilla and \SatOne\add{,} normalized at
    1.1\,$\mu$m\add{,} are shown in Fig.~\ref{fig:spec}.

  \subsection{Spectrum of Camilla\label{sec:spec:107}}
    \indent We combine the near-infrared spectrum we acquired at NASA IRTF
    (Section~\ref{sec:obs:spex}) with the visible
    spectrum from SMASS
    \citep{2002-Icarus-158-BusII,2002-Icarus-158-BusI}
    and analyze them with \add{the} 
    \texttt{M4AST}\footnote{\href{http://m4ast.imcce.fr/}{http://m4ast.imcce.fr/}}
    \citep[Modeling for Asteroids, ][]{2012-AA-544-Popescu} suite of
    Web tools to determine asteroid taxonomic classification,
    mineralogy, and most-likely meteorite analog.
    From this longer wavelength range, we found Camilla to be an Xk-type
    asteroid \citep[using Bus-DeMeo taxonomic scheme,
          Fig.~\ref{fig:spec},][]{2009-Icarus-202-DeMeo}. 
    The low albedo of Camilla
    \citep[0.059\,$\pm$\,0.005, taken as the average of the
      estimates by][]{PDSSBN-TRIAD, 2002-AJ-123-Tedesco-a,
      2010-AJ-140-Ryan, 2011-PASJ-63-Usui,
      2011-ApJ-741-Masiero}\add{,} 
    hints at a P-type classification, using the 
    \citet{1989-AsteroidsII-Tedesco} scheme. 

    \indent Although the best spectral match is formally found
    for an Enstatite Chondrite EH5 meteorite (Queen
    Alexandra Range, Antarctica origin, maximum size of 10\,$\mu$m),
    the low albedo of Camilla argues for a different type of analog
    material.
    The composition of P-type asteroids is indeed \rem{hard} \add{difficult}, if not
    impossible, to infer from their visible and near-infrared spectra
    owing to the lack of absorption bands. 

    \indent Recently, 
    \citet{2015-ApJ-806-Vernazza} have shown
    that anhydrous chondritic porous interplanetary dust particules
    (IDPs) were 
    likely to originate from D- and P-types asteroids, based on 
    spectroscopic observations in the mid-infrared of outer\add{-}belt
    D- and P-type asteroids, including Camilla. The mixture of 
    olivine-rich and pyroxene-rich IDPs they used was compatible with
    the visible and near-infrared spectrum of Camilla.
    As such, the surface of Camilla, and more generally of D- and
    P-types, is very similar to that of
    comets, as already reported by \citet{2006-Icarus-182-Emery}
    from the spectroscopy of Jupiter Trojans in the mid-infrared,
    revealing the presence of anhydrous silicates.


  \subsection{Spectrum of \SatOne\label{sec:spec:S1}}

    \indent As visible \rem{on} \add{in} Fig.~\ref{fig:spec}\add{,}
    the spectrum of \SatOne~is \rem{very} similar to that of Camilla. 
    No significant difference in slope nor absorption band
    can be detected. This implies that the two components
    are spectraly identical from 0.95 to 1.65\,$\mu$m\add{,} within
    the precision of our measurements.
    Such a similarity between the components of multiple systems have
    already been reported for 
    several other main-belt asteroids:
    (22) Kalliope \citep{2009-Icarus-204-Laver},
    (90) Antiope \citep{2009-MPS-44-Polishook,
      2011-Icarus-213-Marchis},  
    (130) Elektra \citep{2016-ApJ-820-Yang}, and
    (379) Huenna \citep{2011-Icarus-212-DeMeo}. 

    \indent Such spectral similarity, together with the main
    characteristics of the orbit (prograde, equatorial, circular)
    supports an origin of these satellites, here for \SatOne~in
    particular, by impact and reaccumulation of material in orbit
    \citep[see][for a review]{2015-AsteroidsIV-Margot}.
    Formation by \rem{rotation} \add{rotational} fission is \rem{here} unlikely owing to the
    rotation period of Camilla (4.84\,h).

\section{Physical properties\label{sec:phys}}

  \subsection{Data processing\label{sec:phys:data}}

    \indent We used the optical lightcurves \rem{at their face
      value}\add{without modification}, only 
    converting their heterogeneous formats from many observers to the
    usual lightcurve inversion format \citep{2010-AA-513-Durech}.
    \add{For the occultation observations, the} \rem{The} location of observers\add{,} together with their timings of
    the disappearance and the reappearance of the star\add{,} were
    converted into chords on the plane of the sky, using the recipes from
    \citet{1999-IMCCE-Berthier}.
    Finally, the 2-D profile of the apparent disk of Camilla was measured
    on the AO images, deconvolved using the
    \mistral~algorithm \citep{2000-PhD-Fusco, 2004-JOSAA-21-Mugnier},
    the reliability of which has been demonstrated elsewhere
    \citep{2006-JGR-111-Witasse}, 
    using the wavelet transform described in 
    \citet{2008-AA-478-Carry, 2010-AA-523-Carry}.

  \subsection{3-D shape modeling with \KOALA\label{sec:shape:koala}}

    \indent We used the multi-data inversion algorithm Knitted Occultation,
    Adaptive-optics, and Lightcurve Analysis (\KOALA), which
    determines the set of rotation period, spin-vector coordinates, and
    3-D shape that provide the best fit to all observations
    simultaneously \citep{2010-Icarus-205-Carry-a}.

    \indent The \KOALA~algorithm minimizes the total
    $\chi^2 = \chi^2_{LC} + w_{AO}\ \chi^2_{AO}+ w_{Occ}\ \chi^2_{Occ}$ 
    that \add{is composed of} \rem{composes} the individual contributions from light curves (LC), 
    profiles from disk-resolved images (AO), and occultation chords
    (Occ). Adaptive optics and occultation data are weighted with
    respect to the lightcurves with parameters $w_{AO}$  and $w_{Occ}$,
    respectively. 
    \add{Within each type of data, all the epochs are weighted
      uniformly.} 
    The optimum values of these weights can be
    objectively obtained
    following the approach of \cite{2011-IPI-5-Kaasalainen}.

    \indent This method has been spectacularly validated by the images
    taken by the 
    OSIRIS camera on-board the ESA Rosetta mission during its flyby of
    the asteroid (21) Lutetia \citep{2011-Science-334-Sierks}.
    Before the encounter, the spin and 3-D shape of Lutetia had been
    determined with \KOALA, using lightcurves and AO images
    \citep{2010-AA-523-Carry, 2010-AA-523-Drummond}. 
    A comparison of the pre-flyby solution with the OSIRIS images
    showed that the \rem{spin-vector} \add{spin vector}
    \rem{coordinates were} \add{was} accurate to within
    2\degr\rem{,} and the diameter to within 2\%. The \add{RMS} residual\rem{s} in the
    \add{surface topography} \rem{topographic profiles} between the \KOALA~predictions and the OSIRIS
    images \rem{were} \add{was} only 2\,km, for a
    98\,km-diameter asteroid \citep{2012-PSS-66-Carry}. 

  \subsection{Spin and 3-D shape of Camilla\label{sec:shape:107}}

\begin{figure*}[ht]
  \includegraphics[width=\textwidth]{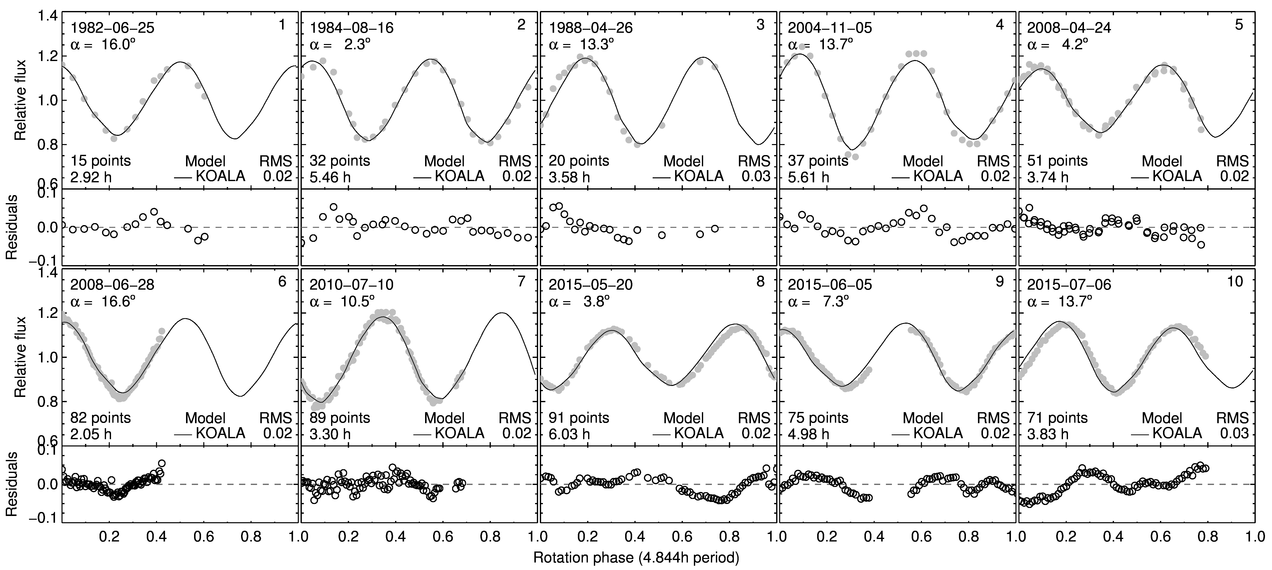}
  \caption[Optical lightcurves of Camilla]{
    \add{Examples of optical lightcurves of Camilla.
    For each epoch, the upper panel presents the observed photometry
    (grey spheres) compared with synthetic lightcurves generated with
    the shape model (black lines). The lower panel shows the residuals
    between the observed and synthetic flux.
    The observing date, number of points, duration
    of the lightcurve (in hours), phase angle ($\alpha$), 
    and RMS residuals between the
    observations and the synthetic lightcurves are displayed.
    In most cases, measurement uncertainties are not provided by the
    observers but can be estimated from the spread of measurements.
    See Fig.~\ref{app:lc} for the entire data set.}
  }
  \label{fig:lc}
\end{figure*}

\begin{figure*}[ht]
  \centering
  \includegraphics[width=.9\textwidth]{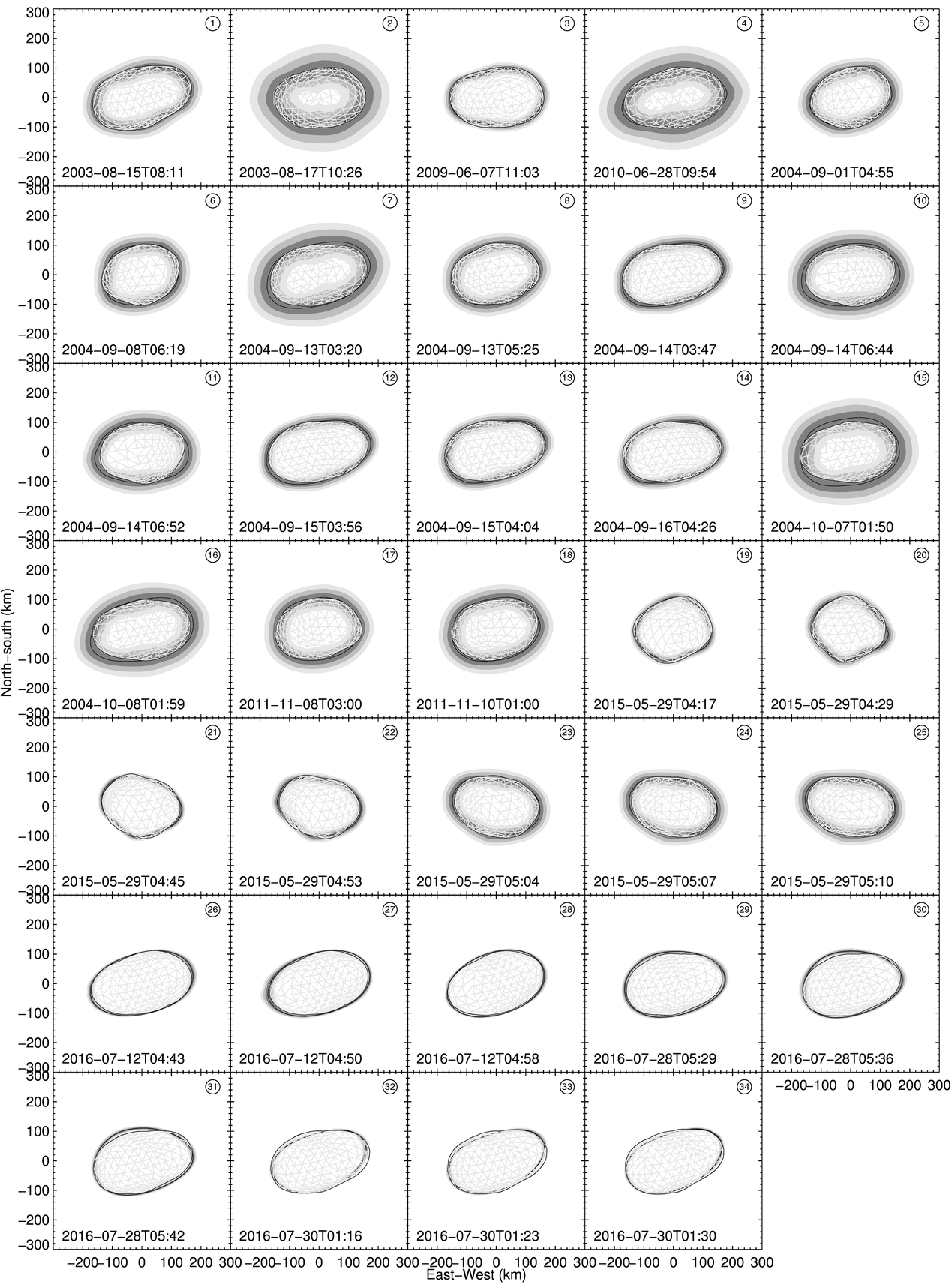}
  \caption[Disk-resolved profiles of Camilla]{
    \rem{Profiles of Camilla from disk-resolved images,}
    \add{All 34 profiles of Camilla from disk-resolved images,
      compared with the \add{projection of the shape model on the
        plane of the sky}. 
      On each panel, corresponding to a different epoch, 
      the grey shaded areas correspond to the 1-2-3\,$\sigma$
      confidence intervals of each profile, while the shape model is
      represented by the wired mesh.}
  }
  \label{fig:prof}
\end{figure*}

\begin{figure*}[ht]
  \includegraphics[width=\textwidth]{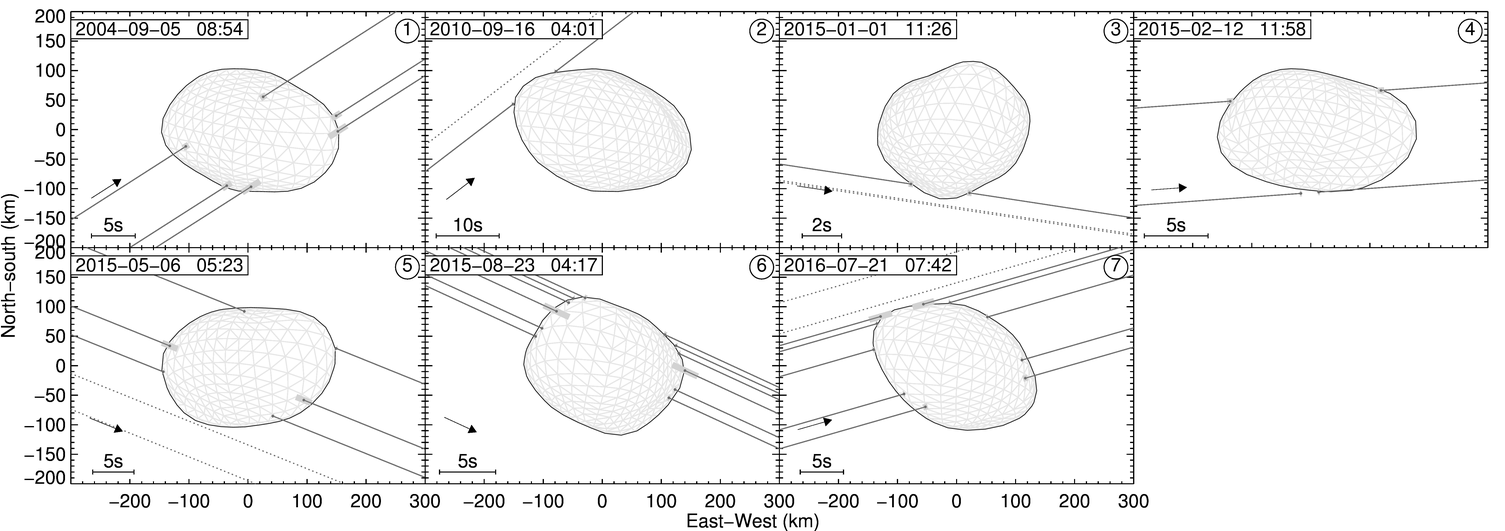}
  \caption[Stellar occultations by Camilla]{
    The seven stellar occultations by Camilla, compared with the \add{shape
    model projected on the 
    plane of the sky for the times of the
    occultations.}
    \add{The observer of northern chord in the first occultation, presenting a
      clear mismatch with the shape model, reported the presence of
      thin cirrus that may explain the discrepancy.}
  }
  \label{fig:occ}
\end{figure*}

    \indent We used
    \numb{\nbLC} optical lightcurves,
    \numb{34} profiles from disk-resolved imaging, and
    \numb{7} stellar occultation events to reconstruct the spin axis and 3-D
    shape of Camilla.
    The model fits \rem{very} well the entire data set, with mean residuals of
    only
    \numb{\rmsLC}\,mag for the lightcurves (\add{Figs.~\ref{fig:lc}, \ref{app:lc}}),
    \numb{\rmsAO}\,pixel for the images (Fig.~\ref{fig:prof}), and
    \numb{\rmsOcc}\,s for the stellar occultations
    (Fig.~\ref{fig:occ}).
    \add{There are small local departures of the shape model from the
      stellar occultation chords that can be due to local topography
      not modeled with our low-resolution shape model.} 

%
%
\begin{figure}[ht]
  \includegraphics[width=.5\textwidth]{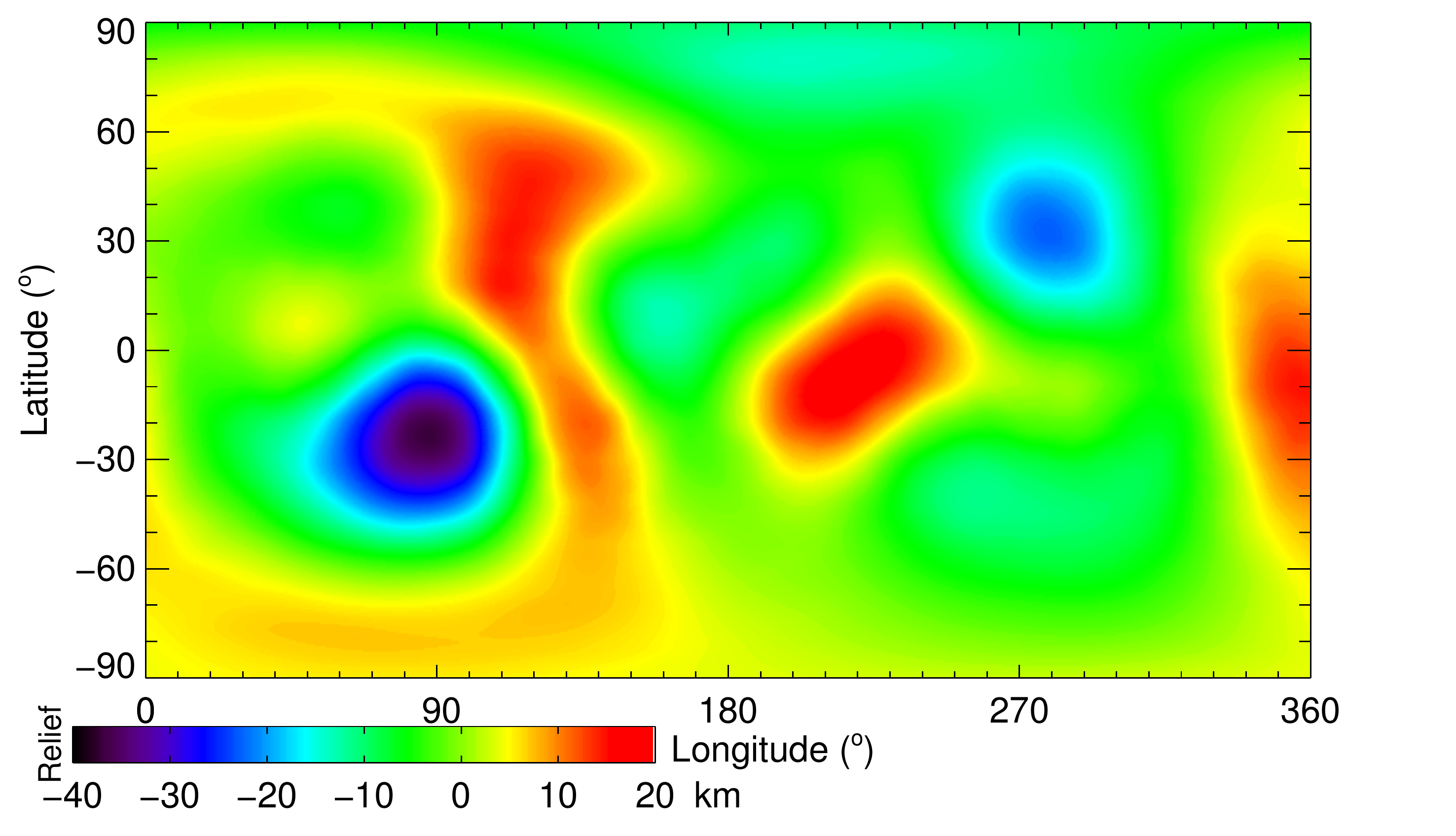}
  \caption[Topography of Camilla]{
    Topographic map of Camilla, with respect to its reference
    ellipsoid (Table~\ref{tab:koala}). The main features are the two
    deep and circular basins located at
    (87\degr,-23\degr)
    and
    (278\degr,+33\degr).
  }
  \label{fig:topo}
\end{figure}

%
    \indent The rotation period and coordinates of the spin axis
    (Table~\ref{tab:koala}) agree very well with previous results from 
    lightcurve-only inversion and convex shape modeling 
    \citep{2003-Icarus-164-Torppa, 2011-Icarus-214-Durech,
      2016-AA-586-Hanus}, as well as 
    models obtained by combining lightcurves and smaller
    subsets of \add{the} present AO data \citep[respectively 3 and 21 epochs,
      see][]{2013-Icarus-226-Hanus, 2017-AA-601-Hanus}.
    The shape of Camilla is far from a sphere, with a strong
    ellipsoidal elongation along the equator (\add{a/b} axes ratio of
    \numb{1.37\,$\pm$\,0.12}, see Table~\ref{tab:koala}).
    Departures from the ellipsoid are, however, limited, and \add{mainly} consist
    \rem{mainly} in two large circular basins, \rem{possibly} reminiscent of
    impact craters (Fig.~\ref{fig:topo}).

%
%
%
    \indent The \add{spherical-}volume-equivalent diameter of Camilla is found to be
    \numb{\Diam\,$\pm$\,\dDiam}\,km (3 $\sigma$), in \rem{perfect} \add{excellent} agreement with the recent
    determination by \citet{2017-AA-601-Hanus} based on a similar data set.
    Both estimates are high compared to diameter estimates from
    infrared observations with IRAS, AKARI, or WISE
    \citep[][see Table~\ref{tab:diam}]{PDSSBN-IRAS,
      2010-AJ-140-Ryan, 2011-PASJ-63-Usui, 2011-ApJ-741-Masiero}.
    However, diameter determinations by 
    mid-IR radiometry are based on  disk-integrated fluxes.
    In the case of highly elongated targets like Camilla, 
    \add{the projected area is often smaller than the average area as
      shown in Table~\ref{tab:IRdiam}. Averaging
    disk-integrated fluxes may thus underestimate the average
    diameter.} 

    \indent The agreement of the 3-D models by 
    \citet{2017-AA-601-Hanus} and developed here with \rem{both 
    the} \add{lightcurves,} disk-resolved images\add{,} and \rem{the} stellar occultation timings,
    providing direct size measurements, indeed
    argues for Camilla being larger than previously thought.
    The corresponding volume is 
    \numb{8.5\,$\pm$\,1.2 $\cdot 10^{6}$}\,km$^3$. The uncertainty
    on the volume matches closely that of the diameter
    ($\delta V / V \approx \delta D / D$) in the case of 3-D shape
    modeling, as shown by \citet{2012-AA-543-Kaasalainen}, 
    because it derives from the uncertainty on the radius of each
    vertex, which are correlated (unlike in the case of scaling a
    sphere).

\begin{table}[ht]
\begin{center}
  \caption[Parameters of the shape model of Camilla]{Sidereal rotation period,
    spin-vector coordinates (longitude $\lambda$, latitude $\beta$ in
    ECJ2000; and right ascension $\alpha$, declination $\delta$ in EQJ2000),
    \add{spherical-}volume-equivalent diameter (D),
    volume (V), 
    diameters along the principal axis of inertia (a, b, c), and
    axes ratio of
    Camilla
    obtained with KOALA.
    All uncertainties are reported at 3 $\sigma$.
    \label{tab:koala}
  }
  \begin{tabular}{llll}
    \hline\hline
    Parameter & Value & Unc. & Unit \\
    \hline
    Period    & 4.843927 & 4.10$^{-5}$ & hour \\
    $\lambda$ & 68.0 & 9.0 & deg. \\
    $\beta$   & 58.3 & 7.0 & deg. \\
    $\alpha$  & 35.8 & 9.0 & deg. \\
    $\delta$  & 76.1 & 7.0 & deg. \\
    T$_0$     & 2444636.00 & & \\
    \hline
    D       & \Diam & \dDiam & km \\
    V       & 8.55 $\cdot 10^{6}$ & 1.21 $\cdot 10^{6}$ & km$^3$ \\
    a       & 340 & \dDiam & km\\
    b       & 249 & \dDiam & km\\
    c       & 197 & \dDiam & km\\
    a/b     & 1.37 & 0.12 & \\
    b/c     & 1.26 & 0.12 & \\
    \hline
  \end{tabular}
\end{center}
\end{table}

\begin{figure*}[ht]
  \centering
  \includegraphics[width=.85\textwidth]{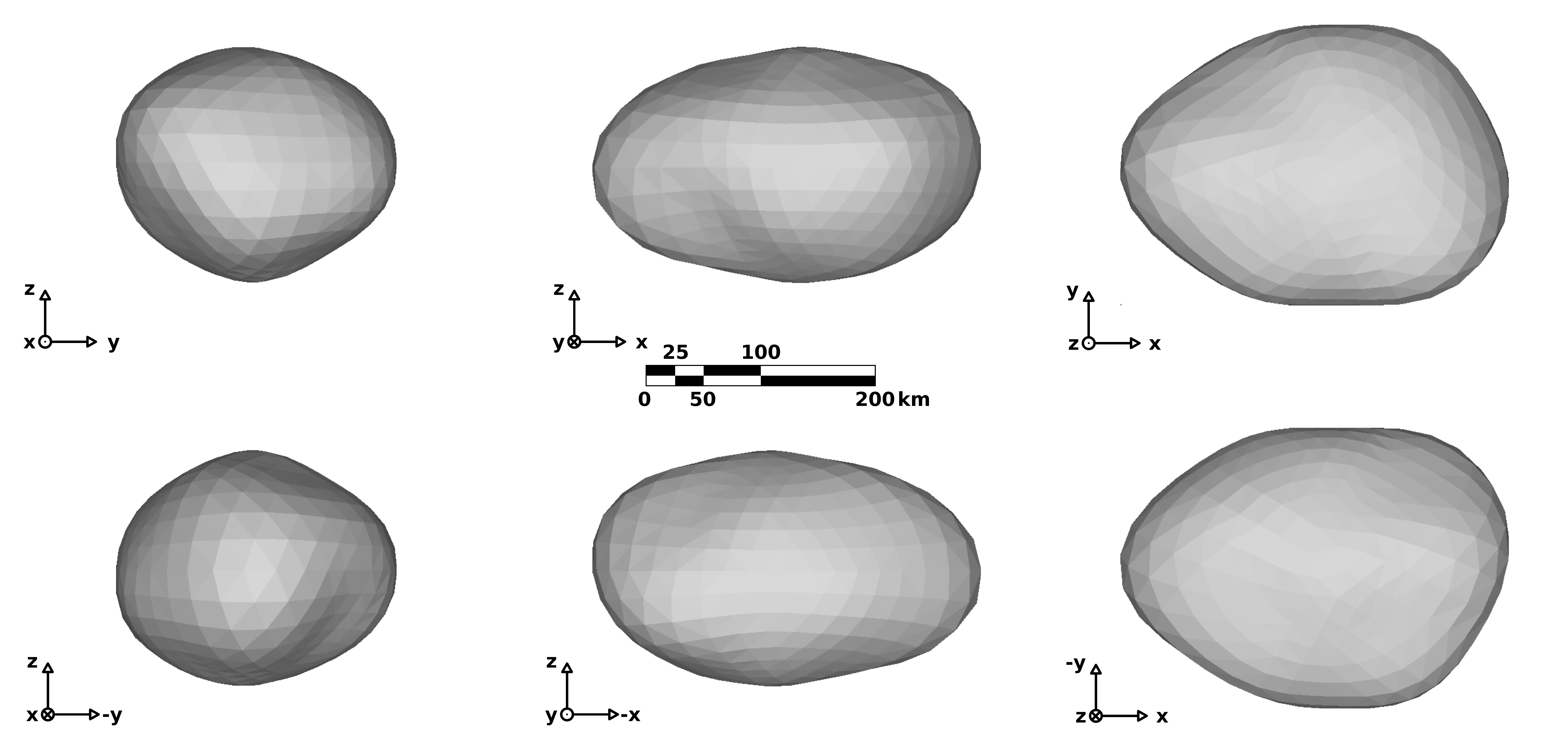}
  \caption[Shape model of Camilla]{
    \add{Views of the shape model along its principal axes (the x,y,z
      axes in the plot are aligned with the principal moment of
      inertia of the model).}
  }
  \label{fig:3d}
\end{figure*}

  \subsection{Diameter of \SatOne\label{sec:shape:S1}}

\begin{figure}[ht]
  \includegraphics[width=.5\textwidth]{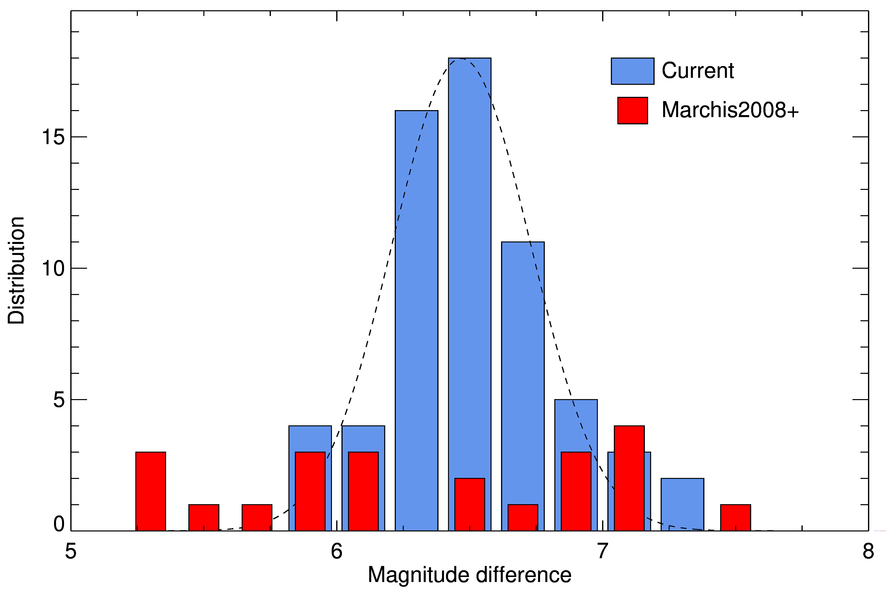}
  \caption[Magnitude of \SatOne~with respect to Camilla]{
    Distribution of the magnitude difference\add{s} between Camilla and its
    largest satellite \SatOne, compared with previous report from
    \citet{2008-Icarus-196-Marchis}.
    The dashed black line represent\add{s} the normal distribution fit to our
    results, with a mean and standard deviation of \numb{6.51\,$\pm$\,0.27}.
  }
    \label{fig:dmag}
  \end{figure}

    \indent We list in Table~\ref{tab:genoid1} and display in 
    Fig.~\ref{fig:dmag} the \numb{65} measured brightness
    difference\add{s} with an uncertainty lower than \numb{1} magnitude
    between Camilla and its largest satellite \SatOne. 
    We found a normal distribution of measurement, as expected from
    photon noise, and measure an average magnitude difference of
    $\Delta m$\,=\,\numb{6.51\,$\pm$\,0.27}, similar to the value of 
    \numb{6.31\,$\pm$\,0.68} reported by 
    \citet{2008-Icarus-196-Marchis} on \numb{22} epochs.

    \indent Using the diameter of 
    \numb{\Diam\,$\pm$\,\dDiam}\,km for Camilla (Sect.~\ref{sec:shape:107})
    and assuming \SatOne~has the same albedo
    as Camilla itself (supported by their spectral similarity, see
    Section~\ref{sec:spec:S1}), this magnitude difference
    implies a size of
    \numb{12.7\,$\pm$\,3.5}\,km for \SatOne, smaller than previously
    reported.

\begin{figure}[ht]
  \includegraphics[width=.5\textwidth]{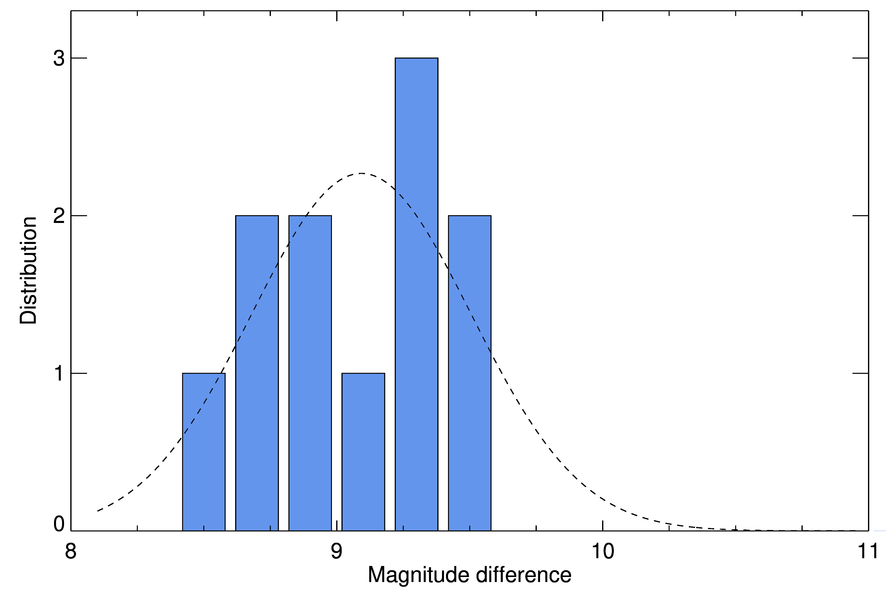}
  \caption[Magnitude of \SatTwo~with respect to Camilla]{
    Distribution of the magnitude difference\add{s} between Camilla and its
    second satellite \SatTwo. 
    The dashed black line represent\add{s} the normal distribution fit to our
    results, with a mean and standard deviation of \numb{9.0\,$\pm$\,0.3}.
  }
    \label{fig:dmag2}
  \end{figure}

  \subsection{Diameter of \SatTwo\label{sec:shape:S2}}
  
    \indent We list in Table~\ref{tab:genoid2} and display in 
    Fig.~\ref{fig:dmag2} the \numb{11} measured brightness
    difference\add{s} between Camilla and its smaller satellite \SatTwo.
    We measure an average magnitude difference of
    $\Delta m$\,=\,\numb{9.0\,$\pm$\,0.3}
    \citep[already reported upon discovery, see][]{2016-IAUC-Marsset}.

    \indent Using the diameter of 
    \numb{\Diam\,$\pm$\,\dDiam}\,km for Camilla (Sect.~\ref{sec:shape:107})
    and assuming \SatTwo~has the same albedo
    as Camilla itself as we did for \SatOne, this magnitude difference
    implies a size of
    \numb{4.0\,$\pm$\,1.2}\,km for \SatTwo.

\section{Discussion\label{sec:discuss}}

  \subsection{Internal Structure\label{sec:dens}}

    \indent Using the mass derived from the study of the \add{dynamics
      of the satellites} and the volume from the 3-D shape modeling, we 
    infer a density of 
    \numb{\dens}\,kg$\cdot$m$^{-3}$ (3 $\sigma$ uncertainty),
    in agreement with previous reports by 
    \citet{2008-Icarus-196-Marchis} and 
    \citet{2017-AA-601-Hanus}.
    This highlights how critically the density relies on accurate
    volume estimates: the summary of previous diameter
    determinations (Table~\ref{tab:diam}),
    mainly based on indirect techniques, 
    leads to a density of 
    \numb{1,750\,$\pm$\,1,400}\,kg$\cdot$m$^{-3}$
    \citep[3 $\sigma$ uncertainty,][]{2012-PSS-73-Carry}.  

    \indent 
      \add{The low density found here is comparable to that
        of (87) Sylvia, a P-type of similar size, also
        orbiting in the Cybele region  
     \citep{2014-Icarus-239-Berthier}, and the D-/P-type 
     Jupiter Trojans (617) Patroclus and (624) Hektor \citep{2010-Icarus-205-Mueller,
        2014-ApJ-783-Marchis,2015-AJ-149-113-Buie}.
      As mentioned above (\ref{sec:spec:107}), the most-likely analog
      material for this type of asteroids are IDPs
      \citep{2015-ApJ-806-Vernazza}.
      There is no measurement of IDP density in the laboratory. However, 
      a density of 3,000$\cdot$m$^{-3}$ for the
      silicate phase was reported by the StarDust mission
      \citep{2006-Science-314-Brownlee}. 
      Because these silicates are mixed with organic carbonaceous
      particles ($\approx$2,200$\cdot$m$^{-3}$),
      the density of the bulk material is likely of
      $\approx$2,600$\cdot$m$^{-3}$
      \citep{2000-EMP-82-Greenberg,2016-Nature-530-Paetzold}.
      A macroporosity of \numb{\poro}\% would thus be required to explain
      the density of Camilla, i.e., half of
      its volume would be occupied by voids.
      Because the pressure inside Camilla reaches 10$^5$\,Pa less
      than 15\,km from its surface (90\% of the radius), it is
      unlikely that its structure can sustain such large voids.
      While silicate grains crush at 10$^7$\,Pa, larger structures
      will not resist pressure significantly 
      smaller, as the compressive strength decreases as the power -1/2
      of the size \citep{1967-JRM-4-Lundborg, 2002-AsteroidsIII-4.2-Britt}.
    }

\begin{figure}[ht]
  \includegraphics[width=.5\textwidth]{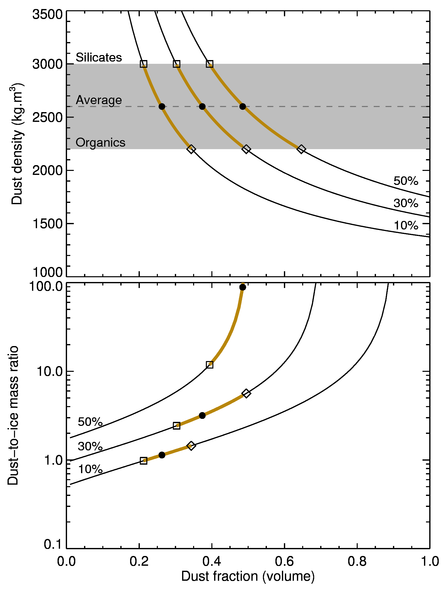}
  \caption[Dust and ice fractions in Camilla]{
    \textbf{Top:} Dust density as function of its volumetric fraction
    for different porosities (10, 30, 50\%).
    The expected range from pure
    organics to pure silicates is represented in shaded gray.
    Expected range is highlighted in gold.
    \textbf{Bottom:} Dust-to-ice mass ratios as function of the
    volumetric fraction of dust. 
  }
  \label{fig:dustice}
\end{figure}

    \indent \add{
      An alternate explanation to the low density of Camilla may be
      that it contains large amounts of water ice.
      An absorption band due to hydration at 3\,$\mu$m was indeed reported by
      \citet{2012-Icarus-219-Takir}, whose shape is similar to those of
      the nearby (24) Themis and (65) Cybele and interpreted as water
      frost coating on surface grains \citep{2010-Nature-464-Campins,
        2011-AA-525-Licandro}.
      Because water ice sublimates on airless
      surfaces at the heliocentric distance of Themis, Camilla, and Cybele, 
      the ice on the surface must be replenishment from sub-surface
      reservoir(s) \citep{2010-Nature-464-Rivkin}, 
      as it occurs on (1) Ceres \citep{1992-Icarus-98-AHearn, 
        2011-AJ-142-Rousselot,
        2014-Nature-505-Kueppers, 
        2016-Science-353-Combe}. 
    }

    \indent \add{
      We thus investigate the possible range of dust-to-ice mass ratio\add{s} 
      as function of macroporosity in Camilla (Fig.~\ref{fig:dustice}).
      As expected, the porosity decreases
      with higher ice content and reaches \numb{10-30\%} for dust-to-ice mass
      ratios of \numb{1--6}.
      Therefore, the volume occupied by dust, ice, and voids would
      be
      \numb{33\,$\pm$\,10\%},
      \numb{47\,$\pm$\,19\%}, and
      \numb{20\,$\pm$\,10\%} respectively, the latter being preferentially
      found in the 
      outer-most volume of the asteroid body.
    }

    \indent \add{To test this, we compute the
      gravitational potential quadrupole
      \numb{$J_2$\,=\,0.042\,$\pm$\,0.004}
      of the 3-D shape model (Sect.~\ref{sec:shape:107})
      under the assumption of a homogeneous interior
      using the method of  
      \citet{1996-Icarus-124-Dobrovolskis}.
      Because the orbit of \SatOne~fits \numb{\ObsOne}~astrometric
      positions over 15 years to 
      measurement accuracy under the assumption of a 
      null $J_2$ (\ref{sec:dyn:S1}), the mass distribution in Camilla
      must be more concentrated at the center, with a denser
        \textsl{core}, than suggested by its
      shape. Similar internal structure has already been suggested 
      for (87) Sylvia and (624) Hektor by 
      \citet{2014-Icarus-239-Berthier} and
      \citet{2014-ApJ-783-Marchis}. 
      Considering a core of pure silicate, and an outer shell of
      porous ice matching the masses above, the core radius would be
      \numb{87\,$\pm$\,8}\,km or
      \numb{68\,$\pm$\,7}\% of the radius of Camilla.
      Additional observations of \SatTwo~to determine precisely its
      orbit are now required to test further the internal structure of
      Camilla.
    }

  \subsection{Future characterization of Camilla triple system\label{sec:dyn:futur}}

    \indent \add{Owing to the large magnitude difference between
      Camilla and its satellites (6.5 and 9 mag.), 
      constraining the size and orbit of the satellites by photometric 
      observations of mutual event
      \citep[eclipses and occultations, see,
        e.g.,][]{2009-Icarus-200-Scheirich,
        2015-Icarus-248-Carry} is not feasible. 
      Observation of \SatTwo~will therefore rely on direct imaging
      such as presented here, or stellar occultations which can moreover
      provide a direct measurement of the diameter of the satellites.
      To this effect, we list in Table~\ref{tab:nextocc} a selection of stellar
      occultations that will occur in the next three years.
    } 

    \indent \add{
      Similarly to our work on (87) Sylvia which led to the
      observation of a stellar occultation by its satellite Romulus
      \citep{2014-Icarus-239-Berthier}, we will continuously update
      the occultation path of Camilla and of its satellites, for
      these events.
      The precision of such predictions will benefit from each
      successive data release of the ESA Gaia astrometry catalogs
      \citep{2007-AA-474-Tanga, 2016-AA-595-Prusti,
        2017-AA-607-Spoto}, that will reduce the uncertainty on the path of
      Camilla itself to a few kilometers.
      The uncertainty on the occultation path of the satellites will
      then mostly derive from the uncertainty on their orbital
      parameters, and we provide them in Table~\ref{tab:nextocc}.
      The orbit of
      \SatTwo~being little constrained, the uncertainty on its
      position for upcoming occultations is very large.
      Initial improvement must thus rely on direct
      imaging of the system.
    }

\section{Summary}

  \indent In the present study, we have acquired and compiled
  optical lightcurves, stellar occultations, visible and near-infrared
  spectra, and high-contrast and
  high-angular-resolution images and spectro-images from the Hubble
  Space Telescope and large ground-based telescopes (Keck, Gemini,
  VLT) equipped with adaptive-optics-fed cameras.

  \indent Using \numb{\ObsOne} positions spanning \numb{15} years, 
  we study the dynamics of the largest satellite, \SatOne, and
  determine its orbit around Camilla to be circular, equatorial, and
  prograde. The residuals between our dynamical solution and the
  observations are \numb{\rmsOne}\,mas, corresponding to a sub-pixel
  accuracy. Using \numb{11} positions of the second, smaller\add{,} 
  satellite \SatTwo~\add{that} we discovered in 2015, we determine a preliminary
  orbit, marginally \rem{tilted} \add{inclined} from that of \SatOne~and more eccentric. 
  Predictions of the relative position of the satellite with
  respect to Camilla, critical for planning stellar occultations for
  instance, are available to the community through our VO service
  \texttt{Miriade}\,\footnote{\href{http://vo.imcce.fr/}{http://vo.imcce.fr/}}
  \citep{2008-ACM-Berthier}.

  \indent From the visible and near-infrared spectrum of Camilla, we
  classify it as an Xk-type asteroid, in the Bus-DeMeo
  taxonomy \citep{2009-Icarus-202-DeMeo}. Considering its low albedo, it would be classified as a
  P-type in older taxonomic schemes such as Tedesco's \citep{1989-AsteroidsII-Tedesco}.
  Using VLT/SPHERE integral-field spectrograph, we measure the
  near-infrared spectrum of the largest satellite, \SatOne, and
  compare it with Camilla. No significant difference\add{s} are found.
  This\add{,} together with its orbital parameters\add{,} argue for a formation of
  the satellite by
  excavation from impact, re-accumulation of ejecta in orbit, and
  circularization by tides. 

  \indent Using optical lightcurves, profiles from disk-resolved imaging, and
  stellar occultation events, we determine the spin-vector coordinates and 3-D
  shape of Camilla. The model fits \rem{very} well each data set, and
  \add{we find a \add{spherical-}volume-equivalent diameter of
    \numb{\Diam\,$\pm$\,\dDiam\,km}.
    By combining the mass from the dynamics with the volume of the shape
    model, we find a density of \numb{\dens}\,kg$\cdot$m$^{-3}$.
    Considering Camilla's most likely analog material are IDPs, this
    implies a macroporosity of \numb{\poro}\%, likely too high to be sustained.
    By considering a mixture of ice and silicate, the macroporosity
    could be in the range 
    \numb{10--30\%} for a dust-to-ice mass ratio of \numb{1--6}, the denser
    material being concentrated toward the center as suggested by the
    dynamics of the system.
  }

\section*{Acknowledgements}

  \indent Based on observations collected at the European Organisation for Astronomical
  Research in the Southern Hemisphere under ESO programmes
  \href{http://archive.eso.org/wdb/wdb/eso/sched_rep_arc/query?progid=71.C-0669}{071.C-0669} (PI Merline),
  \href{http://archive.eso.org/wdb/wdb/eso/sched_rep_arc/query?progid=73.C-0062}{073.C-0062} \&
  \href{http://archive.eso.org/wdb/wdb/eso/sched_rep_arc/query?progid=74.C-0052}{074.C-0052} (PI Marchis),
  \href{http://archive.eso.org/wdb/wdb/eso/sched_rep_arc/query?progid=88.C-0528}{088.C-0528} (PI Rojo),
  \href{http://archive.eso.org/wdb/wdb/eso/sched_rep_arc/query?progid=95.C-0217}{095.C-0217} \&
  \href{http://archive.eso.org/wdb/wdb/eso/sched_rep_arc/query?progid=297.C-5034}{297.C-5034} (PI Marsset).

  \indent Some of the data presented herein were obtained at the W.M. Keck
  Observatory, which is operated as a scientific partnership among the
  California Institute of Technology, the University of California and the
  National Aeronautics and Space Administration. The Observatory was made
  possible by the generous financial support of the W.M. Keck
  Foundation.
  
  \indent This research has made use of the Keck Observatory Archive (KOA),
  which is operated by the W. M. Keck Observatory and the NASA
  Exoplanet Science Institute (NExScI), under contract with the
  National Aeronautics and Space Administration.

  \indent \add{Some of these observations were acquired
    under grants from the National Science Foundation
    and NASA to Merline (PI).}

  \indent The authors wish to recognize and acknowledge the very significant
  cultural role and reverence that the summit of Mauna Kea has always
  had within the indigenous Hawaiian community.  We are most fortunate
  to have the opportunity to conduct observations from this mountain. 

  \indent Based on observations obtained at the Gemini Observatory, which is operated by
  the Association of Universities for Research in Astronomy, Inc., under a
  cooperative agreement with the NSF on behalf of the Gemini partnership: the
  National Science Foundation (United States), the National Research Council
  (Canada), CONICYT (Chile), Ministerio de Ciencia, Tecnolog\'{i}a e
  Innovaci{\'o}n Productiva (Argentina), and Minist{\'e}rio da Ci{\^e}ncia,
  Tecnologia e Inova{\c c}{\~a}o (Brazil). 

  \indent \add{We wish to acknowledge the support of NASA
  Contract NAS5-26555 and STScI grant GO-05583.01 to
  Alex Storrs (PI).} 

  \indent Visiting Astronomer at the Infrared Telescope Facility, which is operated by
  the University of Hawaii under contract NNH14CK55B with the National
  Aeronautics and Space Administration. 
  
  \indent We thank the AGORA association which administrates the
  60\,cm telescope at Les Makes observatory, under a financial agreement
  with Paris Observatory. Thanks to A. Peyrot, J.-P. Teng for local
  support, and A. Klotz for helping with the robotizing.

  \indent \add{Thanks to all the amateurs worldwide who 
    regularly observe asteroid lightcurves and stellar
    occultations. Many co-authors of this study are amateurs who
    observed Camilla, and provided crucial data.}

  \indent We thank J. {\v D}urech for providing his implementation of
  \citet{1996-Icarus-124-Dobrovolskis} method.

  \indent The authors acknowledge the use of the Virtual Observatory
  tools
  \texttt{Miriade}\,\footnote{\href{http://vo.imcce.fr/}{http://vo.imcce.fr/}}
  \citep{2008-ACM-Berthier},
  \texttt{MP$^3$C}\,\footnote{\href{https://mp3c.oca.eu}{https://mp3c.oca.eu}}
  \citep{2017-ACM-Delbo}, 
  \texttt{TOPCAT}\,\footnote{\href{http://www.star.bristol.ac.uk/~mbt/topcat/}{http://www.star.bristol.ac.uk/~mbt/topcat/}},
  and
  \texttt{STILTS}\,\footnote{\href{http://www.star.bristol.ac.uk/~mbt/stilts/}{http://www.star.bristol.ac.uk/~mbt/stilts/}}
  \citep{2005-ASPC-Taylor}.
  This research used the facilities of the Canadian Astronomy Data Centre
  operated by the National Research Council of Canada with the support of the
  Canadian Space Agency \citep{2012-PASP-124-Gwyn}.

 \bibliographystyle{elsarticle-harv} 
 \bibliography{biblio}

\appendix
  \renewcommand{\thefigure}{\Alph{section}.\arabic{figure}}
  \renewcommand{\thetable}{\Alph{section}.\arabic{table}}

\section{Details on the observing data sets}
  \setcounter{figure}{0}
  \setcounter{table}{0}

  \indent We provide here the details for each
  lightcurve (Table~\ref{tab:lc}),
  disk-resolved image (Table~\ref{tab:ao}), and
  stellar occultation (Table~\ref{tab:occ}), as well
  as the astrometry and photometry of
  \SatOne~(Table~\ref{tab:genoid1}) and
  \SatTwo~(Table~\ref{tab:genoid2}).

\onecolumn
\begin{center}
  \begin{longtable}{rcrrrlrcl}
    \caption{
      Date, duration ($\mathcal{L}$, in hours), number of points ($\mathcal{N}_p$), phase angle ($\alpha$), 
      filter, residual (against the shape model), IAU code, and observers, for each 
      lightcurve. \label{tab:lc}
    }\\

    \hline\hline
    & Date & \multicolumn{1}{c}{$\mathcal{L}$} & \multicolumn{1}{c}{$\mathcal{N}_p$} &
    \multicolumn{1}{c}{$\alpha$} & \multicolumn{1}{c}{Filter} & \multicolumn{1}{c}{RMS} &
    \multicolumn{1}{c}{IAU} & \multicolumn{1}{c}{Observers} \\
    && \multicolumn{1}{c}{(h)} && \multicolumn{1}{c}{(\degr)}&& \multicolumn{1}{c}{(mag)} \\
    \hline
    \endfirsthead

    \multicolumn{9}{c}{{\tablename\ \thetable{} -- continued from previous page}} \\ 
    \hline\hline
    & Date & \multicolumn{1}{c}{$\mathcal{L}$} & \multicolumn{1}{c}{$\mathcal{N}_p$} &
    \multicolumn{1}{c}{$\alpha$} & \multicolumn{1}{c}{Filter} & \multicolumn{1}{c}{RMS} &
    \multicolumn{1}{c}{IAU} & \multicolumn{1}{c}{Observers} \\
    && \multicolumn{1}{c}{(h)} && \multicolumn{1}{c}{(\degr)}&& \multicolumn{1}{c}{(mag)} \\
    \hline
    \endhead

    \hline \multicolumn{9}{r}{{Continued on next page}} \\ \hline
    \endfoot

    \hline
    \endlastfoot

      1 & 1981-02-01 &  4.0 &   5 &   2.9 & V     &  0.012 & 654 & \citet{1989-Icarus-81-Harris}          \\
      2 & 1981-02-02 &  6.2 &   9 &   2.8 & V     &  0.023 & 654 & \citet{1989-Icarus-81-Harris}          \\
      3 & 1981-02-04 &  7.7 &  10 &   2.7 & V     &  0.018 & 654 & \citet{1989-Icarus-81-Harris}          \\
      4 & 1981-02-05 &  5.6 &  14 &   2.8 & V     &  0.016 & 654 & \citet{1989-Icarus-81-Harris}          \\
      5 & 1982-01-09 &  2.4 &  11 &  16.4 & V     &  0.021 & 695 & \citet{1987-Icarus-70-Weidenschilling}     \\
      6 & 1982-01-15 &  4.4 &   8 &  16.6 & V     &  0.047 & 695 & \citet{1987-Icarus-70-Weidenschilling}     \\
      7 & 1982-05-20 &  4.5 &  19 &  10.6 & V     &  0.028 & 695 & \citet{1987-Icarus-70-Weidenschilling}     \\
      8 & 1982-06-23 &  4.6 &   6 &  15.8 & V     &  0.030 & 695 & \citet{1987-Icarus-70-Weidenschilling}     \\
      9 & 1982-06-24 &  2.1 &   8 &  15.9 & V     &  0.037 & 695 & \citet{1987-Icarus-70-Weidenschilling}     \\
     10 & 1982-06-25 &  2.9 &  15 &  16.0 & V     &  0.020 & 695 & \citet{1987-Icarus-70-Weidenschilling}     \\
     11 & 1983-03-27 &  2.0 &  10 &  15.5 & V     &  0.048 & 695 & \citet{1987-Icarus-70-Weidenschilling}     \\
     12 & 1983-03-29 &  4.3 &   5 &  15.4 & V     &  0.070 & 695 & \citet{1987-Icarus-70-Weidenschilling}     \\
     13 & 1983-05-24 &  4.8 &  35 &   7.6 & V     &  0.027 & 695 & \citet{1987-Icarus-70-Weidenschilling}     \\
     14 & 1983-07-03 &  4.7 &  23 &   6.1 & V     &  0.021 & 695 & \citet{1987-Icarus-70-Weidenschilling}     \\
     15 & 1984-06-07 &  2.2 &  11 &  14.8 & V     &  0.026 & 695 & \citet{1987-Icarus-70-Weidenschilling}     \\
     16 & 1984-06-10 &  4.6 &  10 &  14.5 & V     &  0.019 & 695 & \citet{1987-Icarus-70-Weidenschilling}     \\
     17 & 1984-07-05 &  3.0 &  15 &  10.7 & V     &  0.047 & 695 & \citet{1987-Icarus-70-Weidenschilling}     \\
     18 & 1984-08-16 &  5.5 &  32 &   2.3 & V     &  0.024 & 809 & \citet{1987-Icarus-69-DiMartino}         \\
     19 & 1985-10-20 &  4.6 &  21 &   3.0 & V     &  0.019 & 695 & \citet{1987-Icarus-70-Weidenschilling}     \\
     20 & 1987-02-06 &  2.7 &  20 &  13.6 & V     &  0.018 & 695 & \citet{1990-Icarus-86-Weidenschilling}     \\
     21 & 1987-02-07 &  4.7 &  17 &  13.8 & V     &  0.019 & 695 & \citet{1990-Icarus-86-Weidenschilling}     \\
     22 & 1988-04-25 &  4.7 &  15 &  13.1 & V     &  0.023 & 695 & \citet{1990-Icarus-86-Weidenschilling}     \\
     23 & 1988-04-26 &  3.6 &  20 &  13.3 & V     &  0.027 & 695 & \citet{1990-Icarus-86-Weidenschilling}     \\
     24 & 1988-04-29 &  2.9 &  16 &  13.9 & V     &  0.032 & 695 & \citet{1990-Icarus-86-Weidenschilling}     \\
     25 & 2004-09-19 &  6.8 &  15 &   2.0 & C     &  0.013 & A14 & L. Bernasconi             \\
     26 & 2004-11-05 &  5.6 &  37 &  13.7 & C     &  0.025 & A14 & L. Bernasconi             \\
     27 & 2008-03-07 &  2.0 &  20 &  11.5 & clear &  0.060 & 950 & SuperWASP - J. Grice                      \\
     28 & 2008-03-13 &  5.2 &  76 &  10.7 & clear &  0.029 & 950 & SuperWASP - J. Grice                      \\
     29 & 2008-03-17 &  3.0 &  53 &   9.4 & clear &  0.028 & 950 & SuperWASP - J. Grice                      \\
     30 & 2008-03-18 &  5.3 & 105 &   8.7 & clear &  0.036 & 950 & SuperWASP - J. Grice                      \\
     31 & 2008-03-29 &  3.3 &  29 &   5.5 & clear &  0.028 & 950 & SuperWASP - J. Grice                      \\
     32 & 2008-03-29 &  4.0 &  40 &   5.5 & clear &  0.033 & 950 & SuperWASP - J. Grice                      \\
     33 & 2008-03-30 &  4.5 &  22 &   5.2 & clear &  0.021 & 950 & SuperWASP - J. Grice                      \\
     34 & 2008-04-01 &  5.0 &  61 &   4.6 & clear &  0.025 & 950 & SuperWASP - J. Grice                      \\
     35 & 2008-04-13 &  4.8 &  69 &   2.2 & clear &  0.041 & 950 & SuperWASP - J. Grice                      \\
     36 & 2008-04-13 &  0.7 &  14 &   2.2 & clear &  0.011 & 950 & SuperWASP - J. Grice                      \\
     37 & 2008-04-16 &  4.9 &  65 &   2.2 & clear &  0.033 & 950 & SuperWASP - J. Grice                      \\
     38 & 2008-04-16 &  4.1 & 128 &   2.4 & clear &  0.027 & 950 & SuperWASP - J. Grice                      \\
     39 & 2008-04-22 &  4.2 &  58 &   3.7 & clear &  0.040 & 950 & SuperWASP - J. Grice                      \\
     40 & 2008-04-22 &  5.3 &  77 &   3.7 & clear &  0.029 & 950 & SuperWASP - J. Grice                      \\
     41 & 2008-04-23 &  3.9 &  44 &   4.0 & clear &  0.030 & 950 & SuperWASP - J. Grice                      \\
     42 & 2008-04-23 &  3.3 &  36 &   4.0 & clear &  0.023 & 950 & SuperWASP - J. Grice                      \\
     43 & 2008-04-24 &  3.2 &  44 &   4.2 & clear &  0.019 & 950 & SuperWASP - J. Grice                      \\
     44 & 2008-04-24 &  3.7 &  51 &   4.2 & clear &  0.020 & 950 & SuperWASP - J. Grice                      \\
     45 & 2008-04-27 &  2.4 &  32 &   5.1 & clear &  0.021 & 950 & SuperWASP - J. Grice                      \\
     46 & 2008-04-28 &  3.3 &  52 &   5.4 & clear &  0.033 & 950 & SuperWASP - J. Grice                      \\
     47 & 2008-04-29 &  5.2 &  61 &   5.7 & clear &  0.029 & 950 & SuperWASP - J. Grice                      \\
     48 & 2008-05-04 &  1.4 &  19 &   7.1 & clear &  0.025 & 950 & SuperWASP - J. Grice                      \\
     49 & 2008-05-09 &  3.8 &  47 &   7.7 & clear &  0.028 & 950 & SuperWASP - J. Grice                      \\
     50 & 2008-05-10 &  3.2 &  93 &   8.5 & clear &  0.026 & 950 & SuperWASP - J. Grice                      \\
     51 & 2008-05-10 &  4.6 & 109 &   8.8 & clear &  0.028 & 950 & SuperWASP - J. Grice                      \\
     52 & 2008-05-12 &  3.6 &  66 &   9.0 & clear &  0.022 & 950 & SuperWASP - J. Grice                      \\
     53 & 2008-05-12 &  3.9 &  81 &   9.3 & clear &  0.048 & 950 & SuperWASP - J. Grice                      \\
     54 & 2008-05-13 &  4.5 &  67 &   9.6 & clear &  0.036 & 950 & SuperWASP - J. Grice                      \\
     55 & 2008-05-19 &  2.4 & 119 &  10.8 & clear &  0.048 & 950 & SuperWASP - J. Grice                      \\
     56 & 2008-05-20 &  3.8 &  69 &  11.3 & clear &  0.048 & 950 & SuperWASP - J. Grice                      \\
     57 & 2008-05-31 &  1.5 &  44 &  13.5 & C     &  0.030 & 181 & \citet{2009-MPBu-Polishook}                      \\
     58 & 2008-06-05 &  1.1 &  13 &  14.3 & clear &  0.043 & 950 & SuperWASP - J. Grice                      \\
     59 & 2008-06-06 &  4.5 &  19 &  14.7 & clear &  0.032 & 950 & SuperWASP - J. Grice                      \\
     60 & 2008-06-06 &  1.2 &  20 &  14.9 & clear &  0.049 & 950 & SuperWASP - J. Grice                      \\
     61 & 2008-06-10 &  4.8 &  83 &  15.0 & clear &  0.044 & 950 & SuperWASP - J. Grice                      \\
     62 & 2008-06-10 &  2.4 &  46 &  16.3 & clear &  0.039 & 950 & SuperWASP - J. Grice                      \\
     63 & 2008-06-23 &  4.2 &  83 &  16.4 & clear &  0.036 & 950 & SuperWASP - J. Grice                      \\
     64 & 2008-06-24 &  4.1 &  81 &  16.5 & clear &  0.032 & 950 & SuperWASP - J. Grice                      \\
     65 & 2008-06-27 &  1.7 &  63 &  16.6 & C     &  0.019 & 181 & \citet{2009-MPBu-Polishook}                      \\
     66 & 2008-06-28 &  2.0 &  82 &  16.6 & C     &  0.018 & 181 & \citet{2009-MPBu-Polishook}                      \\
     67 & 2010-07-09 &  2.7 &  86 &  10.7 & C     &  0.019 & 615 & J. Montier \& S. Heterier \\
     68 & 2010-07-10 &  3.3 &  89 &  10.5 & C     &  0.019 & 517 & F. Reignier                \\
     69 & 2010-07-10 &  3.9 &  57 &  10.5 & clear &  0.063 & 950 & SuperWASP - J. Grice                      \\
     70 & 2010-07-11 &  3.1 & 140 &  10.3 & C     &  0.020 & 615 & J. Montier \& S. Heterier \\
     71 & 2010-07-11 &  4.1 &  85 &  10.2 & clear &  0.024 & 950 & SuperWASP - J. Grice                      \\
     72 & 2010-07-11 &  4.1 &  86 &  10.0 & clear &  0.031 & 950 & SuperWASP - J. Grice                      \\
     73 & 2010-07-13 &  4.0 &  87 &   9.8 & clear &  0.087 & 950 & SuperWASP - J. Grice                      \\
     74 & 2010-07-14 &  4.3 &  93 &   9.6 & clear &  0.040 & 950 & SuperWASP - J. Grice                      \\
     75 & 2010-07-14 &  4.4 &  96 &   9.4 & clear &  0.143 & 950 & SuperWASP - J. Grice                      \\
     76 & 2010-07-16 &  3.3 &  91 &   8.9 & clear &  0.028 & 950 & SuperWASP - J. Grice                      \\
     77 & 2010-07-18 &  2.7 &  91 &   8.4 & clear &  0.036 & 950 & SuperWASP - J. Grice                      \\
     78 & 2010-07-19 &  3.3 & 106 &   8.2 & clear &  0.043 & 950 & SuperWASP - J. Grice                      \\
     79 & 2010-07-20 &  3.1 & 104 &   7.9 & clear &  0.035 & 950 & SuperWASP - J. Grice                      \\
     80 & 2010-07-23 &  2.4 &  54 &   7.4 & clear &  0.035 & 950 & SuperWASP - J. Grice                      \\
     81 & 2010-07-23 &  5.0 &  97 &   7.2 & clear &  0.052 & 950 & SuperWASP - J. Grice                      \\
     82 & 2010-08-01 &  5.3 & 112 &   4.8 & clear &  0.037 & 950 & SuperWASP - J. Grice                      \\
     83 & 2010-08-03 &  2.7 &  47 &   4.3 & clear &  0.035 & 950 & SuperWASP - J. Grice                      \\
     84 & 2010-08-30 &  5.1 & 109 &   4.7 & clear &  0.038 & 950 & SuperWASP - J. Grice                      \\
     85 & 2010-08-31 &  5.3 & 114 &   5.0 & clear &  0.029 & 950 & SuperWASP - J. Grice                      \\
     86 & 2010-09-02 &  1.6 &  22 &   5.5 & clear &  0.023 & 950 & SuperWASP - J. Grice                      \\
     87 & 2010-09-03 &  5.2 & 111 &   5.8 & clear &  0.030 & 950 & SuperWASP - J. Grice                      \\
     88 & 2010-09-04 &  5.1 & 111 &   6.0 & clear &  0.030 & 950 & SuperWASP - J. Grice                      \\
     89 & 2010-09-05 &  3.4 &  71 &   6.3 & clear &  0.070 & 950 & SuperWASP - J. Grice                      \\
     90 & 2010-09-08 &  5.2 &  73 &   7.1 & clear &  0.029 & 950 & SuperWASP - J. Grice                      \\
     91 & 2010-09-08 &  0.9 &  32 &   7.3 & clear &  0.031 & 950 & SuperWASP - J. Grice                      \\
     92 & 2010-09-09 &  1.8 &  23 &   7.6 & clear &  0.024 & 950 & SuperWASP - J. Grice                      \\
     93 & 2010-09-11 &  5.1 & 103 &   7.8 & clear &  0.033 & 950 & SuperWASP - J. Grice                      \\
     94 & 2010-09-30 &  1.9 &  54 &  12.0 & clear &  0.042 & 950 & SuperWASP - J. Grice                      \\
     95 & 2010-10-01 &  4.0 &  84 &  12.1 & clear &  0.052 & 950 & SuperWASP - J. Grice                      \\
     96 & 2015-04-20 &  3.6 &  70 &   8.2 & R     &  0.028 & 181 & F. Vachier                     \\
     97 & 2015-04-21 &  5.7 & 108 &   7.9 & R     &  0.027 & 181 & F. Vachier                     \\
     98 & 2015-04-23 &  5.5 &  87 &   7.4 & R     &  0.025 & 181 & F. Vachier                     \\
     99 & 2015-04-24 &  6.6 & 118 &   7.2 & R     &  0.023 & 181 & F. Vachier                     \\
    100 & 2015-05-09 &  1.4 &  24 &   3.9 & R     &  0.024 & 181 & F. Vachier                     \\
    101 & 2015-05-11 &  4.9 &  84 &   3.6 & R     &  0.021 & 181 & F. Vachier                     \\
    102 & 2015-05-12 &  5.8 &  44 &   3.6 & R     &  0.029 & 181 & F. Vachier                     \\
    103 & 2015-05-13 &  5.2 &  85 &   3.5 & R     &  0.022 & 181 & F. Vachier                     \\
    104 & 2015-05-17 &  3.8 &  58 &   3.5 & R     &  0.025 & 181 & F. Vachier                     \\
    105 & 2015-05-18 &  5.8 &  89 &   3.6 & R     &  0.018 & 181 & F. Vachier                     \\
    106 & 2015-05-19 &  5.1 &  61 &   3.7 & R     &  0.024 & 181 & F. Vachier                     \\
    107 & 2015-05-20 &  6.0 &  91 &   3.8 & R     &  0.021 & 181 & F. Vachier                     \\
    108 & 2015-05-21 &  5.8 & 106 &   4.0 & R     &  0.021 & 181 & F. Vachier                     \\
    109 & 2015-05-22 &  5.4 &  98 &   4.1 & R     &  0.021 & 181 & F. Vachier                     \\
    110 & 2015-05-23 &  6.3 & 102 &   4.3 & R     &  0.021 & 181 & F. Vachier                     \\
    111 & 2015-05-24 &  1.0 &  14 &   4.5 & R     &  0.024 & 181 & F. Vachier                     \\
    112 & 2015-05-26 &  1.9 &  36 &   4.9 & R     &  0.021 & 181 & F. Vachier                     \\
    113 & 2015-06-03 &  3.6 &  68 &   6.8 & R     &  0.022 & 181 & F. Vachier                     \\
    114 & 2015-06-03 &  5.5 & 251 &   6.8 & V     &  0.026 & 517 & D. Romeuf                      \\
    115 & 2015-06-04 &  4.2 &  76 &   7.0 & R     &  0.025 & 181 & F. Vachier                     \\
    116 & 2015-06-05 &  5.0 &  75 &   7.3 & R     &  0.019 & 181 & F. Vachier                     \\
    117 & 2015-06-05 &  4.9 & 274 &   7.3 & V     &  0.026 & 517 & D. Romeuf                      \\
    118 & 2015-06-09 &  3.2 &  59 &   8.3 & R     &  0.024 & 181 & F. Vachier                     \\
    119 & 2015-06-10 &  3.0 &  38 &   8.5 & R     &  0.021 & 181 & F. Vachier                     \\
    120 & 2015-06-11 &  1.4 &  27 &   8.7 & R     &  0.024 & 181 & F. Vachier                     \\
    121 & 2015-06-17 &  5.4 &  98 &  10.1 & R     &  0.024 & 181 & F. Vachier                     \\
    122 & 2015-06-20 & 28.2 & 376 &  10.7 & R     &  0.023 & 586 & S. Fauvaud                      \\
    123 & 2015-06-22 &  5.8 & 104 &  11.2 & R     &  0.052 & 181 & F. Vachier                     \\
    124 & 2015-06-23 &  2.2 &  40 &  11.4 & R     &  0.036 & 181 & F. Vachier                     \\
    125 & 2015-06-25 &  4.7 &  88 &  11.8 & R     &  0.026 & 181 & F. Vachier                     \\
    126 & 2015-06-26 &  3.8 &  70 &  12.0 & R     &  0.029 & 181 & F. Vachier                     \\
    127 & 2015-07-06 &  3.8 &  71 &  13.7 & R     &  0.029 & 181 & F. Vachier                     \\
\hline
  \end{longtable}
\end{center}
\twocolumn

\onecolumn
\begin{center}
  \begin{longtable}{rccrrrrrrrr}
  \caption{
    Date, mid-observing time (UTC), 
    heliocentric distance ($\Delta$) and range to observer ($r$),
    phase angle ($\alpha$), apparent size ($\Theta$), longitude ($\lambda$) and latitude ($\beta$) 
    of the subsolar and subobserver points (SSP, SEP).
    \label{tab:ao}
  }\\

    \hline\hline
     & Date & UTC & \multicolumn{1}{c}{$\Delta$} & \multicolumn{1}{c}{$r$} & \multicolumn{1}{c}{$\alpha$} &
     \multicolumn{1}{c}{$\Theta$} &
     \multicolumn{1}{c}{SEP$_\lambda$} &
     \multicolumn{1}{c}{SEP$_\beta$} &
     \multicolumn{1}{c}{SSP$_\lambda$} &
     \multicolumn{1}{c}{SSP$_\beta$} \\
    &&& \multicolumn{1}{c}{(AU)} & \multicolumn{1}{c}{(AU)} &
    \multicolumn{1}{c}{(\degr)} & \multicolumn{1}{c}{(\arcsec)} & 
    \multicolumn{1}{c}{(\degr)} &\multicolumn{1}{c}{(\degr)} &\multicolumn{1}{c}{(\degr)} &\multicolumn{1}{c}{(\degr)} \\
    \hline
    \endfirsthead

    \multicolumn{11}{c}{{\tablename\ \thetable{} -- continued from previous page}} \\ 
    \hline\hline
     & Date & UTC & \multicolumn{1}{c}{$\Delta$} & \multicolumn{1}{c}{$r$} & \multicolumn{1}{c}{$\alpha$} &
     \multicolumn{1}{c}{$\Theta$} &
     \multicolumn{1}{c}{SEP$_\lambda$} &
     \multicolumn{1}{c}{SEP$_\beta$} &
     \multicolumn{1}{c}{SSP$_\lambda$} &
     \multicolumn{1}{c}{SSP$_\beta$} \\
    &&& \multicolumn{1}{c}{(AU)} & \multicolumn{1}{c}{(AU)} &
    \multicolumn{1}{c}{(\degr)} & \multicolumn{1}{c}{(\arcsec)} & 
    \multicolumn{1}{c}{(\degr)} &\multicolumn{1}{c}{(\degr)} &\multicolumn{1}{c}{(\degr)} &\multicolumn{1}{c}{(\degr)} \\
    \hline
    \endhead

    \hline \multicolumn{11}{r}{{Continued on next page}} \\ \hline
    \endfoot

    \hline
    \endlastfoot

  1 & 2003-08-15 & 08:35:22 &   3.75 &   2.87 &   8.5 &  0.119 &   46.0 &   12.8 &   54.7 &   12.0 \\
  2 & 2003-08-17 & 10:50:07 &   3.75 &   2.88 &   9.0 &  0.117 &  271.7 &   13.0 &  280.9 &   12.0 \\
  3 & 2009-06-07 & 11:25:55 &   3.68 &   2.71 &   5.0 &  0.124 &  267.1 &   15.9 &  264.3 &   20.1 \\
  4 & 2010-06-28 & 10:19:28 &   3.74 &   3.04 &  12.6 &  0.116 &  231.4 &   -1.9 &  221.2 &    5.7 \\
  5 & 2004-09-01 & 05:17:22 &   3.66 &   2.67 &   3.7 &  0.118 &  120.4 &   -8.5 &  117.3 &   -6.3 \\
  6 & 2004-09-08 & 06:41:20 &   3.65 &   2.65 &   1.5 &  0.119 &  129.5 &   -7.6 &  128.3 &   -6.7 \\
  7 & 2004-09-13 & 03:42:51 &   3.65 &   2.65 &   0.2 &  0.142 &   71.4 &   -7.0 &   71.6 &   -6.9 \\
  8 & 2004-09-13 & 05:47:28 &   3.65 &   2.65 &   0.2 &  0.126 &  277.1 &   -7.0 &  277.2 &   -6.9 \\
  9 & 2004-09-14 & 04:09:30 &   3.65 &   2.65 &   0.4 &  0.138 &   54.6 &   -6.9 &   55.0 &   -7.0 \\
 10 & 2004-09-14 & 07:06:44 &   3.65 &   2.65 &   0.5 &  0.139 &  195.0 &   -6.9 &  195.5 &   -7.0 \\
 11 & 2004-09-14 & 07:14:31 &   3.65 &   2.65 &   0.5 &  0.133 &  185.4 &   -6.9 &  185.9 &   -7.0 \\
 12 & 2004-09-15 & 04:18:34 &   3.65 &   2.65 &   0.7 &  0.140 &   59.6 &   -6.8 &   60.3 &   -7.0 \\
 13 & 2004-09-15 & 04:26:55 &   3.65 &   2.65 &   0.7 &  0.134 &   49.2 &   -6.8 &   49.9 &   -7.0 \\
 14 & 2004-09-16 & 04:48:18 &   3.65 &   2.65 &   1.1 &  0.134 &   38.9 &   -6.6 &   39.9 &   -7.1 \\
 15 & 2004-10-07 & 02:12:52 &   3.64 &   2.72 &   7.4 &  0.141 &  212.4 &   -4.3 &  218.7 &   -8.1 \\
 16 & 2004-10-08 & 02:22:37 &   3.64 &   2.73 &   7.6 &  0.136 &  216.6 &   -4.2 &  223.2 &   -8.1 \\
 17 & 2011-11-08 & 03:21:35 &   3.50 &   2.59 &   7.7 &  0.131 &  283.0 &  -13.0 &  289.4 &  -17.7 \\
 18 & 2011-11-10 & 01:22:04 &   3.50 &   2.61 &   8.2 &  0.134 &  103.6 &  -12.9 &  110.5 &  -17.8 \\
 19 & 2015-05-29 & 04:38:45 &   3.58 &   2.61 &   5.4 &  0.120 &  350.8 &   18.1 &  354.8 &   22.0 \\
 20 & 2015-05-29 & 04:51:26 &   3.58 &   2.61 &   5.4 &  0.118 &  335.1 &   18.1 &  339.1 &   22.0 \\
 21 & 2015-05-29 & 05:07:36 &   3.58 &   2.61 &   5.5 &  0.120 &  315.0 &   18.1 &  319.1 &   22.0 \\
 22 & 2015-05-29 & 05:15:12 &   3.58 &   2.61 &   5.5 &  0.122 &  305.6 &   18.1 &  309.6 &   22.0 \\
 23 & 2015-05-29 & 05:25:54 &   3.58 &   2.61 &   5.5 &  0.125 &  292.4 &   18.1 &  296.4 &   22.0 \\
 24 & 2015-05-29 & 05:28:58 &   3.58 &   2.61 &   5.5 &  0.126 &  288.6 &   18.1 &  292.6 &   22.0 \\
 25 & 2015-05-29 & 05:32:04 &   3.58 &   2.61 &   5.5 &  0.127 &  284.7 &   18.1 &  288.8 &   22.0 \\
 26 & 2016-07-12 & 05:06:10 &   3.72 &   2.72 &   3.4 &  0.140 &  233.1 &   10.4 &  233.3 &   13.8 \\
 27 & 2016-07-12 & 05:13:32 &   3.72 &   2.72 &   3.4 &  0.142 &  224.0 &   10.4 &  224.1 &   13.8 \\
 28 & 2016-07-12 & 05:20:55 &   3.72 &   2.72 &   3.4 &  0.137 &  214.8 &   10.4 &  215.0 &   13.8 \\
 29 & 2016-07-28 & 05:52:47 &   3.72 &   2.74 &   4.8 &  0.139 &   74.3 &   11.8 &   79.0 &   13.2 \\
 30 & 2016-07-28 & 05:59:03 &   3.72 &   2.74 &   4.8 &  0.139 &   66.5 &   11.8 &   71.3 &   13.2 \\
 31 & 2016-07-28 & 06:05:21 &   3.72 &   2.74 &   4.8 &  0.140 &   58.7 &   11.8 &   63.5 &   13.2 \\
 32 & 2016-07-30 & 01:39:02 &   3.72 &   2.75 &   5.2 &  0.128 &   61.0 &   11.9 &   66.3 &   13.1 \\
 33 & 2016-07-30 & 01:46:07 &   3.72 &   2.75 &   5.2 &  0.129 &   52.3 &   11.9 &   57.5 &   13.1 \\
 34 & 2016-07-30 & 01:53:12 &   3.72 &   2.75 &   5.2 &  0.130 &   43.5 &   11.9 &   48.7 &   13.1 \\
  \end{longtable}
\end{center}
\twocolumn

\begin{table*}
  \caption{
    Timing and location of each observer for the stellar occultations
    used in this work.
    \label{tab:obsocc}
  }
\begin{tabular}{llrrrr}
\hline
  \multicolumn{1}{c}{Observer} &
  \multicolumn{1}{c}{Location} &
  \multicolumn{1}{c}{Latitude} &
  \multicolumn{1}{c}{Longitude} &
  \multicolumn{1}{c}{Disappearance} &
  \multicolumn{1}{c}{Reappearance} \\
  \multicolumn{1}{c}{} &
  \multicolumn{1}{c}{} &
  \multicolumn{1}{c}{(\degr)} &
  \multicolumn{1}{c}{(\degr)} &
  \multicolumn{1}{c}{(UT)} &
  \multicolumn{1}{c}{(UT)} \\
\hline
  \noalign{\smallskip}
  \multicolumn{2}{c}{2004 September 5} \\
  Randy Peterson & Cave Creek Desert, AZ & 33.813 & -112.0002 & 8:54:49.1 & 8:55:03.9\\
  Paul Maley/Syd Leach & Fountain Hills, AZ & 33.6278 & -111.8333 & 8:54:50.96 & 8:55:02.70\\
  Scott Donnell & Eastonville, CO & 39.0725 & -104.5778 & 8:54:01.5 & 8:54:12.0\\
  \noalign{\smallskip}
  \multicolumn{2}{c}{2010 September 16} \\
  \noalign{\smallskip}
  Kerry Coughlin & LaPaz, Baja, Mexico & 24.1387 & -110.3296 & Miss & Miss\\
  Roc Fleishmann & Todos Santos, Baja, Mexico & 23.4484 & -110.2261 & 4:01:03.91 & 4:01:12.46\\
  \noalign{\smallskip}
  \multicolumn{2}{c}{2015 January 01} \\
  Andy Scheck & Scaggsville, MD & 40.3511 & -77.0 & 11:26:31.76 & 11:26:34.77\\
  Bob Dunford & Naperville, IL & 41.759 & -88.1167 & Miss & Miss\\
  Chad Ellington & Owings, MD & 38.6906 & -76.6354 & Miss & Miss\\
  \noalign{\smallskip}
  \multicolumn{2}{c}{2015 February 12} \\
  Derek Breit & Morgan Hill, CA & 37.1133 & -121.7028 & 11:58:13.97 & 11:58:23.98\\
  Derek Breit (double star) & Morgan Hill, CA & 37.1133 & -121.7028 & 11:58:11.03 & 11:58:25.48\\
  Ted Blank & Payson, AZ & 34.2257 & -111.2988 & 11:58:56.82 & 11:58:58.55\\
  Chuck McPartlin & Santa Barbara, CA & 34.4567 & -119.795 & Miss & Miss\\
  Tony George & Scottsdale, AZ & 33.7157 & -111.8494 & Miss & Miss\\
  Sam Herchak & Mesa, AZ & 33.3967 & -111.6985 & Miss & Miss\\
  \noalign{\smallskip}
  \multicolumn{2}{c}{2015 May 06} \\
  Dan Caton & Boone, NC & 36.2514 & -81.4122 & 5:23:07.46 & 5:23:19.46\\
  Steve Messner & Bevier, MO & 39.7722 & -92.5243 & 5:24:02.71 & 5:24:20.96\\
  Roger Venable & Elgin, SC & 34.1476 & -80.7502 & 5:22:54.75 & 5:23:09.05\\
  Roger Venable & New Holland, SC & 33.7397 & -81.5201 & Miss & Miss\\
  Roger Venable & Hepzibah, GA & 33.3394 & -82.1542 & Miss & Miss\\
  Chris Anderson & Twin Falls, ID & 42.5839 & -114.4703 & Miss & Miss\\
  Chuck McPartlin & Santa Barbara, CA & 34.4568 & -119.7951 & Miss & Miss\\
  \noalign{\smallskip}
  \multicolumn{2}{c}{2015 August 23} \\
  Steve Preston & Carnation, WA & 47.6437 & -121.9224 & 4:17:35.07 & 4:17:42.89\\
  Andrea Dobson/Larry North & Walla Walla, WA & 46.0044 & -118.8928 & 4:17:49.06 & 4:17:59.13\\
  Tony George & Umatilla, OR & 45.9221 & -119.2983 & 4:17:47.72 & 4:17:58.54\\
  Chad Ellington & Tumwater, WA & 46.9763 & -122.9111 & 4:17:31.85 & 4:17:44.46\\
  Chris Anderson & Twin Falls, ID & 42.5839 & -114.4703 & 4:18:14.41 & 4:18:27.46\\
  David Becker & Grasmere, ID & 42.6733 & -115.8981 & 4:18:10.17 & 4:18:23.25\\
  William Gimple & Greenville, CA & 40.1377 & -120.8667 & Miss & Miss\\
  Charles Arrowsmith & Quincy, CA & 39.9477 & -120.9691 & Miss & Miss\\
  Tom Beard & Reno, NV & 39.3729 & -119.8312 & Miss & Miss\\
  Jerry Bardecker & Gardnerville, NV & 38.8899 & -119.6723 & Miss & Miss\\
  Ted Swift & Davis, CA & 38.5522 & -121.7856 & Miss & Miss\\
  \noalign{\smallskip}
  \multicolumn{2}{c}{2016 July 21} \\
  Derek Breit & Morgan Hill, CA & 37.1133 & -121.7028 & Miss & Miss\\
  Bob Dunford & Naperville, IL & 41.759 & -88.1167 & Miss & Miss\\
  Brad Timerson & Newark, NY & 43.0066 & -77.1185 & Miss & Miss\\
  Kevin Green & Westport, CT & 41.1714 & -73.3278 & Miss & Miss\\
  Steve Conard & Gamber, MD & 39.4692 & -76.9516 & Miss & Miss\\
  Gary Frishkorn & Sykesville, MD & 39.2316 & -76.9929 & 7:41:59.05 & 7:42:04.22\\
  Andy Scheck & Scaggsville, MD & 39.1497 & -76.8871 & 7:41:57.87 & 7:42:06.68\\
  David Dunham/Joan Dunham & Greenbelt, MD & 38.9866 & -76.8694 & 7:41:56.97 & no report\\
  Paul Maley & Clifton, TX & 31.6814 & -97.6744 & 7:43:26.99 & 7:43:40.62\\
  Ned Smith & Trenton, GA & 34.893 & -85.4711 & 7:42:28.72 & 7:42:42.92\\
  Ernie Iverson & Lufkin, TX & 31.3213 & -94.8444 & 7:43:15.48 & 7:43:27.47\\
\hline\end{tabular}
\end{table*}

\begin{table*}
\begin{center}
  \caption{
    Date, number of positive and negative chords (\#$_p$ and \#$_n$), 
    average uncertainty in seconds ($\sigma_s$) and kilometers ($\sigma_k$), and 
    RMS residuals with seconds, kilometers, and expressed in amount of standard deviation.
    \label{tab:occ}
  }
  \begin{tabular}{rccrrrrrrr}

    \hline\hline
     & Date & UT & \multicolumn{1}{c}{\#$_p$} & \multicolumn{1}{c}{\#$_n$} & 
     \multicolumn{1}{c}{$\sigma_s$} &
     \multicolumn{1}{c}{$\sigma_k$} &
     \multicolumn{1}{c}{RMS$_s$} &
     \multicolumn{1}{c}{RMS$_k$} &
     \multicolumn{1}{c}{RMS$_\sigma$} \\
    && (h) &&& \multicolumn{1}{c}{(s)} & \multicolumn{1}{c}{(km)} &
    \multicolumn{1}{c}{(s)} & \multicolumn{1}{c}{(km)} & \multicolumn{1}{c}{($\sigma$)} \\
    \hline
    1   & 2004-09-05 & 08:54 &   3 &   0 &  0.73 & 17.831 &  0.860 & 32.358 &  3.277 \\
    2   & 2010-09-16 & 04:01 &   1 &   1 &  0.05 &  0.267 &  0.066 &  0.995 &  1.318 \\
    3   & 2015-01-01 & 11:26 &   1 &   2 &  0.22 &  2.445 &  0.017 &  0.975 &  0.077 \\
    4   & 2015-02-12 & 11:58 &   2 &   0 &  0.20 &  6.728 &  1.230 & 22.304 &  8.171 \\
    5   & 2015-05-06 & 05:23 &   3 &   4 &  0.33 &  3.831 &  0.389 & 13.324 & 13.219 \\
    6   & 2015-08-23 & 04:17 &   6 &   5 &  0.15 &  4.994 &  0.072 &  8.039 &  2.362 \\
    7   & 2016-07-21 & 07:42 &   5 &   5 &  0.56 &  5.060 &  0.579 & 18.920 &  3.067 \\
    0   & Average    &    -- &   3 &   2 &  0.32 &  5.880 &  0.459 & 13.845 &  4.499 \\
    \hline
  \end{tabular}
\end{center}
\end{table*}

\onecolumn
\begin{center}
  \begin{longtable}{cclllrrrrrrr}
  \caption[Astrometry of \SatOne]{Astrometry of \SatOne.
    Date, mid-observing time (UTC), telescope, camera, filter, 
    astrometry ($X$
    is aligned with Right Ascension, and $Y$ with Declination, and
    $o$ and $c$ indices stand for observed and computed positions),
    and photometry (magnitude difference $\Delta M$ with uncertainty
  $\delta M$). PIs of these observations were: 
  $^{*}$A. Storrs,$^{a}$J.-L. Margot, $^{b}$W. J. Merline,
  $^{c}$L. Sromovsky, $^{d}$F. Marchis, $^{e}$P. Rojo, and $^{f}$M. Marsset. 
    \label{tab:genoid1}
  }\\

    \hline\hline
     Date & UTC & Tel. & Cam. & Filter &
     \multicolumn{1}{c}{$X_o$} &
     \multicolumn{1}{c}{$Y_o$} &
     \multicolumn{1}{c}{$X_{o-c}$} &
     \multicolumn{1}{c}{$Y_{o-c}$} &
     \multicolumn{1}{c}{$\sigma$} &
     \multicolumn{1}{c}{$\Delta M$} &
     \multicolumn{1}{c}{$\delta M$} \\
    &&&&& 
     \multicolumn{1}{c}{(mas)} & \multicolumn{1}{c}{(mas)} &
     \multicolumn{1}{c}{(mas)} & \multicolumn{1}{c}{(mas)} & 
     \multicolumn{1}{c}{(mas)} & 
     \multicolumn{1}{c}{(mag)} &\multicolumn{1}{c}{(mag)}  \\
    \hline
    \endfirsthead

    \multicolumn{11}{c}{{\tablename\ \thetable{} -- continued from previous page}} \\ 
    \hline\hline
     Date & UTC & Tel. & Cam. & Filter &
     \multicolumn{1}{c}{$X_o$} &
     \multicolumn{1}{c}{$Y_o$} &
     \multicolumn{1}{c}{$X_{o-c}$} &
     \multicolumn{1}{c}{$Y_{o-c}$} &
     \multicolumn{1}{c}{$\sigma$} &
     \multicolumn{1}{c}{$\Delta M$} &
     \multicolumn{1}{c}{$\delta M$} \\
    &&&&& 
     \multicolumn{1}{c}{(mas)} & \multicolumn{1}{c}{(mas)} &
     \multicolumn{1}{c}{(mas)} & \multicolumn{1}{c}{(mas)} & 
     \multicolumn{1}{c}{(mas)} & 
     \multicolumn{1}{c}{(mag)} &\multicolumn{1}{c}{(mag)}  \\
    \hline
    \endhead

    \hline \multicolumn{11}{r}{{Continued on next page}} \\ \hline
    \endfoot

    \hline
    \endlastfoot

2001-03-01 & 05:48:13.0 & HST      & ACS$^{*}$      & F439W    & -573 &  -84 &  -22 &    1 &  10.00 &   0.00 &   0.00 \\
2001-03-01 & 06:00:12.9 & HST      & ACS$^{*}$      & F791W    & -565 &  -70 &  -20 &   13 &  10.00 &   0.00 &   0.00 \\
2002-05-08 & 10:46:01.0 & Keck     & NIRC2$^{a}$    & Kp       &  472 & -189 &   -4 &   15 &   9.94 &   6.34 &   1.50 \\
2003-06-06 & 14:03:06.0 & Keck     & NIRC2$^{b}$    & Ks       &  402 & -214 &   -5 &   -8 &   9.94 &   6.53 &   1.18 \\
2003-06-06 & 14:08:23.2 & Keck     & NIRC2$^{b}$    & Ks       &  406 & -213 &    1 &   -8 &   9.94 &   7.18 &   0.45 \\
2003-06-06 & 14:13:30.2 & Keck     & NIRC2$^{b}$    & Ks       &  402 & -218 &   -1 &  -12 &   9.94 &   6.31 &   0.23 \\
2003-07-15 & 07:32:50.4 & VLT      & NACO$^{b}$     & H        & -540 &  216 &    8 &    7 &  27.00 &   6.56 &   0.02 \\
2003-07-15 & 07:37:26.2 & VLT      & NACO$^{b}$     & H        & -536 &  222 &   10 &   12 &  27.00 &   6.34 &   0.14 \\
2003-08-14 & 10:35:08.0 & Keck     & NIRC2$^{a}$    & H        & -183 &  227 &    5 &   -4 &   9.94 &   5.04 &   3.68 \\
2003-08-15 & 08:35:22.2 & Keck     & NIRC2$^{c}$    & Kp       &  554 &  -62 &    5 &   -1 &   9.94 &   6.67 &   0.18 \\
2003-08-15 & 08:39:27.2 & Keck     & NIRC2$^{c}$    & Kp       &  550 &  -66 &    0 &   -5 &   9.94 &   6.62 &   0.31 \\
2003-08-17 & 10:50:08.0 & Keck     & NIRC2$^{b}$    & Kp       & -568 &  146 &    8 &    2 &   9.94 &   6.55 &   1.21 \\
2003-08-17 & 10:53:39.3 & Keck     & NIRC2$^{b}$    & Kp       & -567 &  144 &    9 &    0 &   9.94 &   6.39 &   0.66 \\
2004-09-01 & 05:07:38.3 & VLT      & NACO$^{d}$     & Ks       &  504 & -164 &    4 &    1 &  13.24 &   6.06 &   0.31 \\
2004-09-01 & 05:17:22.2 & VLT      & NACO$^{d}$     & H        &  510 & -165 &    5 &    0 &  13.24 &   6.34 &   0.24 \\
2004-09-01 & 08:06:43.4 & VLT      & NACO$^{d}$     & Ks       &  576 & -169 &    8 &   -1 &  13.24 &   6.98 &   0.43 \\
2004-09-03 & 06:51:57.5 & VLT      & NACO$^{d}$     & Ks       & -623 &  166 &  -21 &    2 &  13.24 &   6.76 &   0.70 \\
2004-09-05 & 04:28:20.2 & VLT      & NACO$^{d}$     & Ks       &  624 & -163 &    6 &    0 &  13.24 &   6.73 &   0.09 \\
2004-09-08 & 06:41:20.1 & VLT      & NACO$^{d}$     & Ks       &  211 & -120 &    4 &   -8 &  13.24 &   6.95 &   0.59 \\
2004-09-11 & 04:34:26.2 & VLT      & NACO$^{d}$     & Ks       & -539 &   87 &   -5 &   -9 &  13.24 &   7.09 &   1.15 \\
2004-09-13 & 03:42:52.5 & VLT      & NACO$^{d}$     & Ks       &  470 &  -75 &    1 &    0 &  13.24 &   7.23 &   1.22 \\
2004-09-13 & 05:47:28.2 & VLT      & NACO$^{d}$     & Ks       &  386 &  -46 &  -15 &    6 &  13.24 &   6.08 &   1.51 \\
2004-09-14 & 04:09:30.3 & VLT      & NACO$^{d}$     & Ks       & -500 &  153 &   -9 &   -1 &  13.24 &   6.59 &   0.14 \\
2004-09-15 & 04:18:34.3 & VLT      & NACO$^{d}$     & Ks       & -321 &   35 &    0 &    4 &  13.24 &   6.28 &   1.20 \\
2004-09-15 & 04:26:56.5 & VLT      & NACO$^{d}$     & H        & -315 &   36 &    0 &    6 &  13.24 &   7.30 &   0.69 \\
2004-10-07 & 02:02:03.0 & VLT      & NACO$^{d}$     & Ks       & -540 &  123 &    4 &   -4 &  13.24 &   8.49 &   1.55 \\
2004-10-08 & 02:22:38.3 & VLT      & NACO$^{d}$     & Ks       &  356 & -106 &    3 &    0 &  13.24 &   8.22 &   2.62 \\
2004-10-08 & 04:47:21.2 & VLT      & NACO$^{d}$     & Ks       &  435 & -125 &    3 &   -1 &  13.24 &   7.07 &   1.05 \\
2004-10-20 & 00:39:22.2 & VLT      & NACO$^{d}$     & Ks       &  553 & -136 &    8 &    0 &  13.24 &   6.55 &   0.19 \\
2004-11-02 & 07:36:13.0 & Gemini   & NIRI$^{b}$     & Kp       & -344 &   88 &   -2 &    0 &  21.90 &   6.49 &   0.62 \\
2004-11-02 & 07:38:36.9 & Gemini   & NIRI$^{b}$     & Kp       & -340 &   90 &    0 &    2 &  21.90 &   6.40 &   0.29 \\
2004-11-05 & 08:09:18.1 & Gemini   & NIRI$^{b}$     & Kp       & -538 &  138 &   -6 &    1 &  21.90 &   5.95 &   0.16 \\
2005-12-21 & 09:05:51.5 & Gemini   & NIRI$^{d}$     & Ks       &  684 &    0 &    8 &   -6 &  21.90 &   6.53 &   0.02 \\
2006-01-01 & 10:17:12.1 & Gemini   & NIRI$^{d}$     & Ks       &  651 &  -35 &    6 &   -7 &  21.90 &   6.71 &   0.17 \\
2006-01-09 & 05:20:11.1 & Gemini   & NIRI$^{d}$     & Ks       &  557 &  116 &   17 &    1 &  21.90 &   5.86 &   0.34 \\
2006-01-16 & 05:16:51.5 & Gemini   & NIRI$^{d}$     & Ks       &  619 &  -17 &   22 &   31 &  21.90 &   5.85 &   0.28 \\
2009-06-07 & 10:29:14.1 & Keck     & NIRC2$^{b}$    & H        &  510 &   54 &    0 &   -1 &   9.94 &   6.56 &   0.44 \\
2009-06-07 & 10:32:18.1 & Keck     & NIRC2$^{b}$    & H        &  511 &   55 &    0 &    0 &   9.94 &   6.49 &   0.50 \\
2009-06-07 & 10:54:08.0 & Keck     & NIRC2$^{b}$    & Kp       &  516 &   55 &   -4 &    7 &   9.94 &   6.23 &   0.44 \\
2009-06-07 & 11:23:04.0 & Keck     & NIRC2$^{b}$    & Kp       &  530 &   44 &   -2 &    5 &   9.94 &   6.56 &   1.07 \\
2009-06-07 & 11:25:56.5 & Keck     & NIRC2$^{b}$    & Kp       &  530 &   39 &   -3 &    0 &   9.94 &   6.66 &   0.17 \\
2009-08-16 & 06:47:02.0 & Keck     & NIRC2$^{d}$    & FeII     &  -36 &  239 &    6 &    0 &   9.94 &   6.99 &   1.19 \\
2010-08-15 & 08:07:02.0 & Gemini   & NIRI$^{d}$     & Kp       & -421 &  182 &    0 &   14 &  21.90 &   5.84 &   0.08 \\
2010-08-15 & 08:16:53.5 & Gemini   & NIRI$^{d}$     & Kp       & -412 &  181 &    4 &   13 &  21.90 &   6.40 &   0.05 \\
2010-08-28 & 08:49:11.1 & Gemini   & NIRI$^{d}$     & Kp       &  379 & -189 &    0 &  -14 &  21.90 &   6.05 &   0.28 \\
2010-08-28 & 08:54:01.0 & Gemini   & NIRI$^{d}$     & Kp       &  378 & -186 &    0 &  -12 &  21.90 &   6.52 &   0.47 \\
2010-09-02 & 06:45:32.3 & Gemini   & NIRI$^{d}$     & Kp       & -588 &  157 &   -3 &    9 &  21.90 &   6.02 &   0.10 \\
2010-10-31 & 05:58:48.4 & Gemini   & NIRI$^{b}$     & Kp       & -271 &   -8 &   27 &   -8 &  21.90 &   6.67 &   0.05 \\
2010-10-31 & 06:03:23.2 & Gemini   & NIRI$^{b}$     & Kp       & -290 &   -2 &   10 &   -3 &  21.90 &   6.87 &   0.06 \\
2011-09-27 & 05:04:41.0 & VLT      & NACO$^{d}$     & H        & -287 &  236 &    5 &   12 &  13.24 &   6.39 &   1.09 \\
2011-09-29 & 05:21:18.0 & VLT      & NACO$^{d}$     & H        &  440 & -217 &   -3 &   -8 &  13.24 &   7.04 &   1.18 \\
2011-11-08 & 03:21:35.3 & VLT      & NACO$^{e}$     & H        & -438 &  -61 &   -6 &    0 &  13.24 &   6.66 &   0.06 \\
2011-11-10 & 01:22:04.0 & VLT      & NACO$^{e}$     & H        &  386 &   93 &    5 &   17 &  13.24 &   7.35 &   0.17 \\
2015-05-29 & 04:38:46.4 & VLT      & SPHERE$^{f}$   & Ks       & -188 &  240 &    3 &    0 &  12.26 &   6.29 &   0.06 \\
2015-05-29 & 04:38:46.4 & VLT      & SPHERE$^{f}$   & YJH      & -184 &  237 &    6 &   -3 &   7.40 &   6.28 &   0.07 \\
2015-05-29 & 04:51:27.2 & VLT      & SPHERE$^{f}$   & YJH      & -176 &  239 &    4 &   -1 &   7.40 &   6.30 &   0.13 \\
2015-05-29 & 04:51:27.2 & VLT      & SPHERE$^{f}$   & Ks       & -180 &  241 &    1 &    0 &  12.26 &   6.26 &   0.09 \\
2015-05-29 & 05:07:36.3 & VLT      & SPHERE$^{f}$   & Ks       & -166 &  245 &    3 &    2 &  12.26 &   6.26 &   0.06 \\
2015-05-29 & 05:07:36.3 & VLT      & SPHERE$^{f}$   & YJH      & -164 &  241 &    5 &   -1 &   7.40 &   6.35 &   0.09 \\
2015-05-29 & 05:15:13.1 & VLT      & SPHERE$^{f}$   & YJH      & -157 &  242 &    6 &    0 &   7.40 &   6.43 &   0.30 \\
2015-05-29 & 05:15:13.1 & VLT      & SPHERE$^{f}$   & Ks       & -158 &  245 &    5 &    2 &  12.26 &   6.35 &   0.17 \\
2015-05-29 & 05:25:55.5 & VLT      & SPHERE$^{f}$   & Ks       & -152 &  245 &    3 &    1 &  12.26 &   6.36 &   0.16 \\
2015-05-29 & 05:28:59.5 & VLT      & SPHERE$^{f}$   & YJH      & -148 &  243 &    4 &    0 &   7.40 &   6.49 &   0.16 \\
2015-05-29 & 05:28:59.5 & VLT      & SPHERE$^{f}$   & Ks       & -150 &  247 &    3 &    2 &  12.26 &   6.34 &   0.08 \\
2015-05-29 & 05:32:04.0 & VLT      & SPHERE$^{f}$   & Ks       & -148 &  245 &    2 &    0 &  12.26 &   6.42 &   0.09 \\
2015-05-29 & 05:32:04.0 & VLT      & SPHERE$^{f}$   & YJH      & -146 &  243 &    4 &   -1 &   7.40 &   6.51 &   0.17 \\
2016-07-02 & 08:47:22.2 & VLT      & SPHERE$^{f}$   & YJH      & -279 &  -90 &   -5 &   -4 &   7.40 &   6.68 &   0.28 \\
2016-07-12 & 05:04:19.4 & VLT      & SPHERE$^{f}$   & YJH      &  601 & -129 &   -9 &    1 &   7.40 &   6.55 &   0.19 \\
2016-07-12 & 05:11:41.7 & VLT      & SPHERE$^{f}$   & YJH      &  601 & -130 &  -10 &    1 &   7.40 &   6.51 &   0.06 \\
2016-07-12 & 05:19:03.9 & VLT      & SPHERE$^{f}$   & YJH      &  601 & -129 &  -10 &    4 &   7.40 &   6.49 &   0.03 \\
2016-07-28 & 05:50:56.0 & VLT      & SPHERE$^{f}$   & YJH      & -208 & -138 &    7 &   -5 &   7.40 &   6.96 &   0.24 \\
2016-07-28 & 05:57:12.3 & VLT      & SPHERE$^{f}$   & YJH      & -212 & -137 &    8 &   -6 &   7.40 &   7.10 &   0.06 \\
2016-07-28 & 06:03:30.1 & VLT      & SPHERE$^{f}$   & YJH      & -216 & -135 &    7 &   -5 &   7.40 &   7.07 &   0.19 \\
2016-07-30 & 01:37:12.1 & VLT      & SPHERE$^{f}$   & YJH      &  192 &  141 &   -1 &    0 &   7.40 &   6.90 &   0.20 \\
2016-07-30 & 01:44:17.2 & VLT      & SPHERE$^{f}$   & YJH      &  194 &  135 &   -4 &   -4 &   7.40 &   6.78 &   0.23 \\
2016-07-30 & 01:51:22.2 & VLT      & SPHERE$^{f}$   & YJH      &  199 &  135 &   -4 &   -2 &   7.40 &   6.65 &   0.44 \\
2016-08-11 & 00:18:43.4 & VLT      & SPHERE$^{f}$   & YJH      &  579 & -159 &   -9 &    6 &   7.40 &   6.18 &   0.07 \\
2016-08-11 & 02:41:34.1 & VLT      & SPHERE$^{f}$   & YJH      &  559 & -189 &   -6 &    5 &   7.40 &   6.54 &   0.23 \\
2016-08-11 & 02:48:41.5 & VLT      & SPHERE$^{f}$   & YJH      &  560 & -189 &   -3 &    6 &   7.40 &   6.43 &   0.08 \\
2016-08-11 & 02:55:50.8 & VLT      & SPHERE$^{f}$   & YJH      &  556 & -194 &   -5 &    2 &   7.40 &   6.43 &   0.07 \\
\hline
&&&&& \multicolumn{3}{c}{Average}                 1 &    0 &   18 &   6.50 &   0.46 \\ 
&&&&& \multicolumn{3}{c}{Standard deviation}      8 &    7 &    7 &   0.28 &   0.61 \\ 
  \end{longtable}
\end{center}

\begin{center}
  \begin{longtable}{cclllrrrrrrr}
  \caption[Astrometry of \SatTwo]{Astrometry of \SatTwo.
    Date, mid-observing time (UTC), telescope, camera, filter, 
    astrometry ($X$
    is aligned with Right Ascension, and $Y$ with Declination, and
    $o$ and $c$ indices stand for observed and computed positions),
    and photometry (magnitude difference $\Delta M$ with uncertainty
  $\delta M$). The PI of these observations was M. Marsset.
    \label{tab:genoid2}
  }\\

    \hline\hline
     Date & UTC & Tel. & Cam. & Filter &
     \multicolumn{1}{c}{$X_o$} &
     \multicolumn{1}{c}{$Y_o$} &
     \multicolumn{1}{c}{$X_{o-c}$} &
     \multicolumn{1}{c}{$Y_{o-c}$} &
     \multicolumn{1}{c}{$\sigma$} &
     \multicolumn{1}{c}{$\Delta M$} &
     \multicolumn{1}{c}{$\delta M$} \\
    &&&&& 
     \multicolumn{1}{c}{(mas)} & \multicolumn{1}{c}{(mas)} &
     \multicolumn{1}{c}{(mas)} & \multicolumn{1}{c}{(mas)} & 
     \multicolumn{1}{c}{(mas)} & 
     \multicolumn{1}{c}{(mag)} &\multicolumn{1}{c}{(mag)}  \\
    \hline
    \endfirsthead

    \multicolumn{11}{c}{{\tablename\ \thetable{} -- continued from previous page}} \\ 
    \hline\hline
     Date & UTC & Tel. & Cam. & Filter &
     \multicolumn{1}{c}{$X_o$} &
     \multicolumn{1}{c}{$Y_o$} &
     \multicolumn{1}{c}{$X_{o-c}$} &
     \multicolumn{1}{c}{$Y_{o-c}$} &
     \multicolumn{1}{c}{$\sigma$} &
     \multicolumn{1}{c}{$\Delta M$} &
     \multicolumn{1}{c}{$\delta M$} \\
    &&&&& 
     \multicolumn{1}{c}{(mas)} & \multicolumn{1}{c}{(mas)} &
     \multicolumn{1}{c}{(mas)} & \multicolumn{1}{c}{(mas)} & 
     \multicolumn{1}{c}{(mas)} & 
     \multicolumn{1}{c}{(mag)} &\multicolumn{1}{c}{(mag)}  \\
    \hline
    \endhead

    \hline \multicolumn{11}{r}{{Continued on next page}} \\ \hline
    \endfoot

    \hline
    \endlastfoot

2015-05-29 & 04:38:46.4 & VLT      & SPHERE   & YJH      &   87 &  140 &   -3 &   10 &   7.40 &   8.95 &   1.40 \\
2015-05-29 & 04:51:27.2 & VLT      & SPHERE   & YJH      &  102 &  141 &    1 &    7 &   7.40 &   8.65 &   0.25 \\
2015-05-29 & 05:07:36.3 & VLT      & SPHERE   & YJH      &  111 &  137 &   -2 &   -1 &   7.40 &   8.43 &   1.53 \\
2015-05-29 & 05:15:13.1 & VLT      & SPHERE   & YJH      &  121 &  142 &    1 &    1 &   7.40 &   8.66 &   0.60 \\
2015-05-29 & 05:32:04.0 & VLT      & SPHERE   & YJH      &  135 &  136 &    2 &   -9 &   7.40 &   8.83 &   1.59 \\
2016-07-12 & 05:04:19.4 & VLT      & SPHERE   & YJH      & -271 &  115 &   10 &    0 &   7.40 &   9.16 &   0.82 \\
2016-07-12 & 05:11:41.7 & VLT      & SPHERE   & YJH      & -275 &  113 &    2 &   -5 &   7.40 &   9.53 &   1.23 \\
2016-07-12 & 05:19:03.9 & VLT      & SPHERE   & YJH      & -272 &  119 &    0 &   -2 &   7.40 &   9.34 &   0.95 \\
2016-07-30 & 01:37:12.1 & VLT      & SPHERE   & YJH      & -295 &  104 &   -2 &    5 &   7.40 &   9.32 &   0.33 \\
2016-07-30 & 01:44:17.2 & VLT      & SPHERE   & YJH      & -295 &  103 &   -6 &    0 &   7.40 &   9.23 &   0.20 \\
2016-07-30 & 01:51:22.2 & VLT      & SPHERE   & YJH      & -288 &  102 &   -3 &   -3 &   7.40 &   9.53 &   1.69 \\
\hline
&&&&& \multicolumn{3}{c}{Average}                 0 &    0 &   10 &   9.05 &   0.96 \\ 
&&&&& \multicolumn{3}{c}{Standard deviation}      4 &    5 &    0 &   0.32 &   0.56 \\ 
  \end{longtable}
\end{center}
\twocolumn

\section{Previous determinations of mass and diameter}
  \setcounter{figure}{0}
  \setcounter{table}{0}

\begin{center}
\begin{table*}[h]
  \caption[Mass estimates of (107) Camilla]{
    The mass estimates ($\mathcal{M}$) of (107) Camilla collected
    \rem{in} \add{from} the literature.
    For each, the 3\,$\sigma$ uncertainty, method, selection flag, and 
    bibliographic reference are reported. The methods are
    \textsc{bin}: Binary, \textsc{defl}: Deflection, \textsc{ephem}: Ephemeris.
    \label{tab:mass} 
..."
  }
  \begin{tabular}{rrlcl}
    \hline\hline
     \multicolumn{1}{c}{\#} & \multicolumn{1}{c}{Mass ($\mathcal{M}$)} &
     \multicolumn{1}{c}{Method} & \multicolumn{1}{c}{Sel.} & \multicolumn{1}{c}{Reference}  \\
    & \multicolumn{1}{c}{(kg)} \\
    \hline
  1 & $(1.12 \pm 0.09) \times 10^{19}$                   & \textsc{bim}  & \ding{51} & \citet{2008-Icarus-196-Marchis}          \\
  2 & $(36.20 \pm 27.72) \times 10^{18}$                 & \textsc{ephem} & \ding{55} & \citet{2010-SciNote-Fienga}              \\
  3 & $3.88_{-3.88}^{+32.70} \times 10^{18}$             & \textsc{defl}  & \ding{51} & \citet{2011-AJ-142-Zielenbach}           \\
  4 & $(39.00 \pm 31.80) \times 10^{18}$                 & \textsc{defl}  & \ding{55} & \citet{2011-AJ-142-Zielenbach}           \\
  5 & $(17.60 \pm 26.07) \times 10^{18}$                 & \textsc{defl}  & \ding{51} & \citet{2011-AJ-142-Zielenbach}           \\
  6 & $2.25_{-2.25}^{+54.00} \times 10^{18}$             & \textsc{defl}  & \ding{51} & \citet{2011-AJ-142-Zielenbach}           \\
  7 & $(27.10 \pm 20.88) \times 10^{18}$                 & \textsc{ephem} & \ding{55} & \citet{2011-DPS-Fienga}                  \\
  8 & $(6.79 \pm 9.00) \times 10^{18}$                   & \textsc{ephem} & \ding{51} & \citet{2012-SciNote-Fienga}              \\
  9 & $(11.10 \pm 5.37) \times 10^{18}$                  & \textsc{defl}  & \ding{51} & \citet{2014-AA-565-Goffin}               \\
 10 & $(16.10 \pm 13.26) \times 10^{18}$                 & \textsc{ephem} & \ding{51} & \citet{2017-BDL-108-Viswanathan}         \\
 11 & $(1.12 \pm 0.01) \times 10^{19}$                   & \textsc{bin} & \ding{51} & This work                        \\
\hline
 & $(1.12 \pm 0.09) \times 10^{19}$ & \multicolumn{2}{c}{Average} \\
   \hline
  \end{tabular}
\end{table*}
\end{center}

\begin{center}
\begin{table*}[h]
  \caption[Diameter estimates of (107) Camilla]{
    The diameter estimates ($\mathcal{D}$) of (107) Camilla collected
    \rem{in} \add{from} the literature.
    For each, the 3\,$\sigma$ uncertainty, method, selection flag, and 
    bibliographic reference are reported. The methods are
    \textsc{im}: Disk-Resolved Imaging, \textsc{adam}/\textsc{koala}: Multidata 3-D
    Modeling, \textsc{lcimg}: 3-D Model scaled with Imaging,
    \textsc{lcocc}: 3-D Model scaled with Occultation, \textsc{neatm}:
    Near-Earth Asteroid Thermal Model, \textsc{stm}: Standard Thermal
    Model, \textsc{tpm}: Thermophysical Model. 
    \label{tab:diam} 
  }
  \begin{tabular}{rrrlcl}
    \hline\hline
     \multicolumn{1}{c}{\#} & \multicolumn{1}{c}{$\mathcal{D}$} & \multicolumn{1}{c}{$\delta \mathcal{D}$} &
     \multicolumn{1}{c}{Method} & \multicolumn{1}{c}{Sel.} & \multicolumn{1}{c}{Reference}  \\
    & \multicolumn{1}{c}{(km)} & \multicolumn{1}{c}{(km)} \\
    \hline
  1 &    213.00 &   63.90 & \textsc{stm}   & \ding{51} & \citet{PDSSBN-TRIAD}                     \\
  2 &    222.62 &   51.30 & \textsc{stm}   & \ding{51} & \citet{PDSSBN-IRAS}                      \\
  3 &    185.00 &   27.00 & \textsc{im}    & \ding{55} & \citet{2006-Icarus-185-Marchis}          \\
  4 &    249.00 &   54.00 & \textsc{neatm} & \ding{51} & \citet{2008-Icarus-196-Marchis}          \\
  5 &    246.00 &   39.00 & \textsc{im}    & \ding{51} & \citet{2008-Icarus-196-Marchis}          \\
  6 &    208.85 &   32.37 & \textsc{stm}   & \ding{51} & \citet{2010-AJ-140-Ryan}                 \\
  7 &    221.10 &   43.11 & \textsc{neatm} & \ding{51} & \citet{2010-AJ-140-Ryan}                 \\
  8 &    214.00 &   84.00 & \textsc{lcocc} & \ding{51} & \citet{2011-Icarus-214-Durech}           \\
  9 &    200.37 &   10.53 & \textsc{stm}   & \ding{55} & \citet{2011-PASJ-63-Usui}                \\
 10 &    219.37 &   17.82 & \textsc{neatm} & \ding{51} & \citet{2011-ApJ-741-Masiero}             \\
 11 &    256.00 &   35.00 & \textsc{neatm} & \ding{51} & \citet{2012-Icarus-221-Marchis}          \\
 12 &    245.00 &   75.00 & \textsc{tpm}   & \ding{51} & \citet{2012-Icarus-221-Marchis}          \\
 13 &    227.00 &   72.00 & \textsc{lcimg} & \ding{51} & \citet{2013-Icarus-226-Hanus}            \\
 14 &    254.00 &   18.00 & \textsc{adam} & \ding{51} & \citet{2017-AA-601-Hanus}                \\
 15 &    254.00 &   36.00 & \textsc{koala} & \ding{51} & This work                        \\
\hline
 &    234.72 &     53.11 & \multicolumn{2}{c}{Average} \\
   \hline
  \end{tabular}
\end{table*}
\end{center}

\begin{figure}[h!]
\centering
\includegraphics[width=\linewidth]{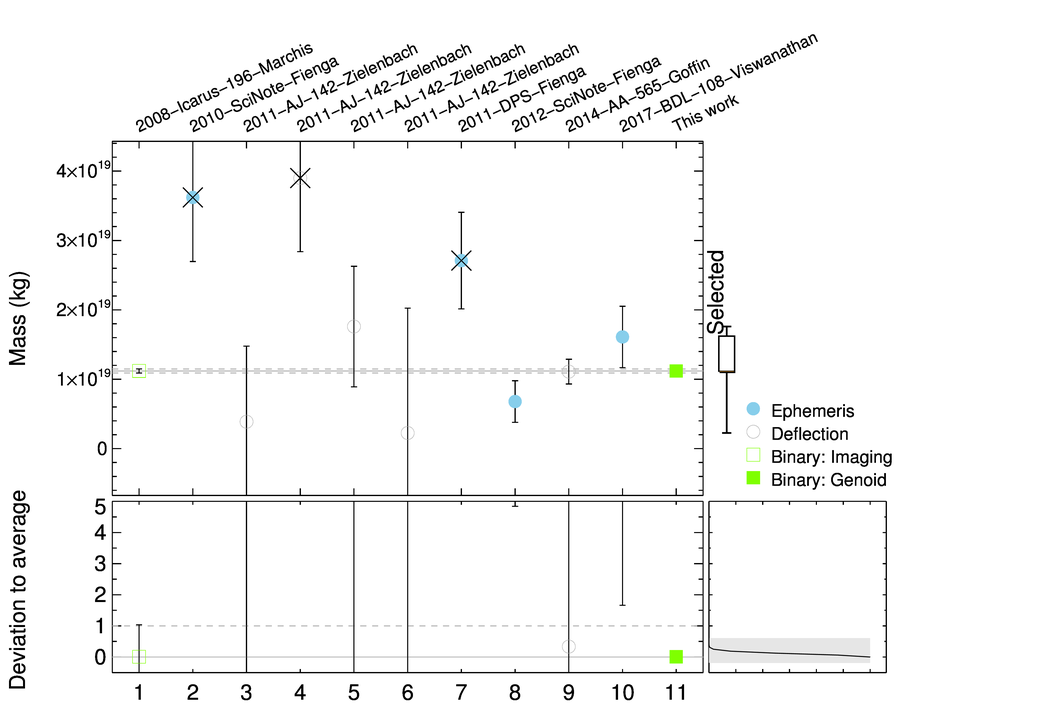}
 \caption[Mass estimates of (107)~Camilla]{Mass estimates of (107)~Camilla gathered from the
   literature, see Table~\ref{tab:mass} for details.} 
\label{fig:mass}
\end{figure}

\begin{figure}[h!]
\centering
\includegraphics[width=\linewidth]{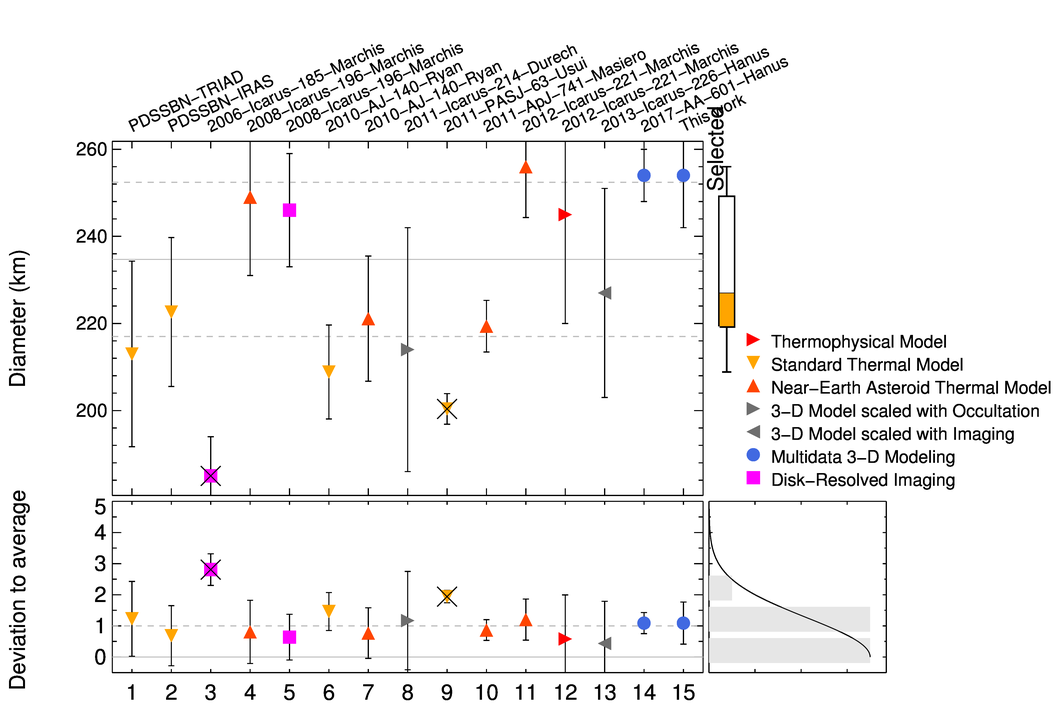}
 \caption[Diameter estimates of (107)~Camilla]{Diameter estimates of (107)~Camilla gathered from the
   literature, see Table~\ref{tab:diam} for details.} 
\label{fig:diam}
\end{figure}

\begin{table}[ht]
\begin{center}
  \caption[Diameter of Camilla as seen from mid-IR satellite]{\add{Spherical-equivalent diameter ($\mathcal{D}_e$) of the shape model of
      Camilla projected on the plane of the sky as seen from 
      IRAS, AKARI, and WISE \cite{PDSSBN-IRAS, 2011-PASJ-63-Usui,
        2011-ApJ-741-Masiero}.
      Owing to the elongated shape of Camilla, the 2-D diameter often
      underestimates the spherical-volume equivalent diameter.
      \label{tab:IRdiam}
  }}
  \begin{tabular}{llr}
    \hline\hline
         & Epoch (UTC) & $\mathcal{D}_e$ \\
    \hline
    IRAS &    1983-03-14T12:55      &      242.1 \\
         &    1983-03-14T14:26      &      265.3 \\
         &    1983-03-22T01:11      &      257.8 \\
         &    1983-03-21T23:28      &      266.5 \\
         &    1983-03-29T20:25      &      266.0 \\
         &    1983-03-30T12:04      &      236.6 \\
         &    1983-09-30T08:21      &      244.5 \\
         &    1983-09-30T10:04      &      263.5 \\
         &    1983-09-30T06:38      &      243.2 \\
         &     Average              &      254.0 \\
         &     Standard deviation   &       12.2 \\
    \hline
    WISE &    2010-05-18T06:50      &      241.8 \\
         &    2010-05-18T10:01      &      228.0 \\
         &    2010-05-18T13:11      &      261.5 \\
         &    2010-05-18T19:32      &      222.3 \\
         &    2010-05-18T21:08      &      255.9 \\
         &    2010-05-18T22:43      &      256.0 \\
         &    2010-05-19T00:18      &      220.7 \\
         &    2010-05-19T01:53      &      259.9 \\
         &    2010-05-19T03:29      &      252.9 \\
         &    2010-05-19T06:39      &      262.7 \\
         &    2010-05-19T09:50      &      220.5 \\
         &    2010-05-19T13:00      &      244.1 \\
         &    2010-11-05T05:36      &      238.3 \\
         &    2010-11-05T15:08      &      248.2 \\
         &    2010-11-05T16:43      &      259.7 \\
         &    2010-11-05T18:18      &      225.1 \\
         &    2010-11-05T18:18      &      225.1 \\
         &    2010-11-05T23:04      &      222.8 \\
         &    2010-11-05T23:04      &      222.8 \\
         &     Average              &      240.4 \\
         &    Standard deviation    &       16.3 \\
    \hline
    AKARI&    2006-11-05T21:59      &      232.9 \\
         &    2006-11-05T23:38      &      236.0 \\
         &    2007-04-29T08:45      &      240.4 \\
         &    2007-04-29T10:24      &      265.1 \\
         &    2007-04-29T20:21      &      264.7 \\
         &     Average              &      247.8 \\
         &    Standard deviation    &       15.8 \\  
    \hline
  \end{tabular}
\end{center}
\end{table}

\clearpage

\section{Fit to the optical lightcurves}
  \setcounter{figure}{0}
  \setcounter{table}{0}


\begin{figure*}[]
  \centering
  \begin{subfigure}{\textwidth}
    \includegraphics[width=.9\linewidth]{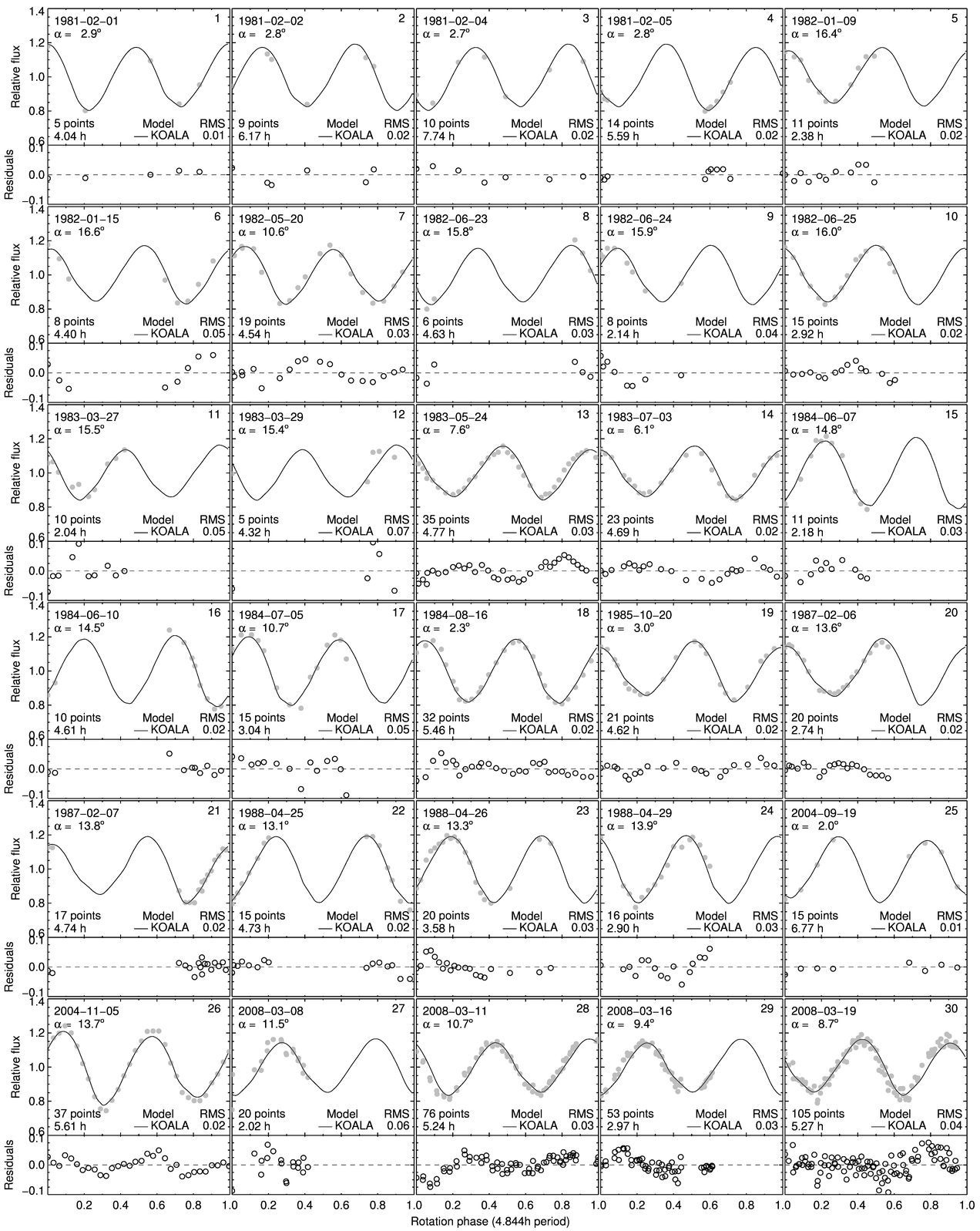}
  \end{subfigure}
  \caption[Optical lightcurves of Camilla]{\add{The optical lightcurves of Camilla (grey spheres),
    compared with the synthetic lightcurves generated with the shape
    model (black lines). 
    On each panel, the observing date, number of points, duration
    of the lightcurve (in hours), and RMS residuals between the
    observations and the synthetic lightcurves from the shape model
    are displayed.
    In many cases, measurement uncertainties are not provided by the
    observers but can be estimated from the spread of measurements. }
  \label{app:lc}}
\end{figure*}
\clearpage
\begin{figure*}[]
  \centering
  \ContinuedFloat
  \captionsetup{list=off}
  \begin{subfigure}{.9\textwidth}
    \includegraphics[width=\linewidth]{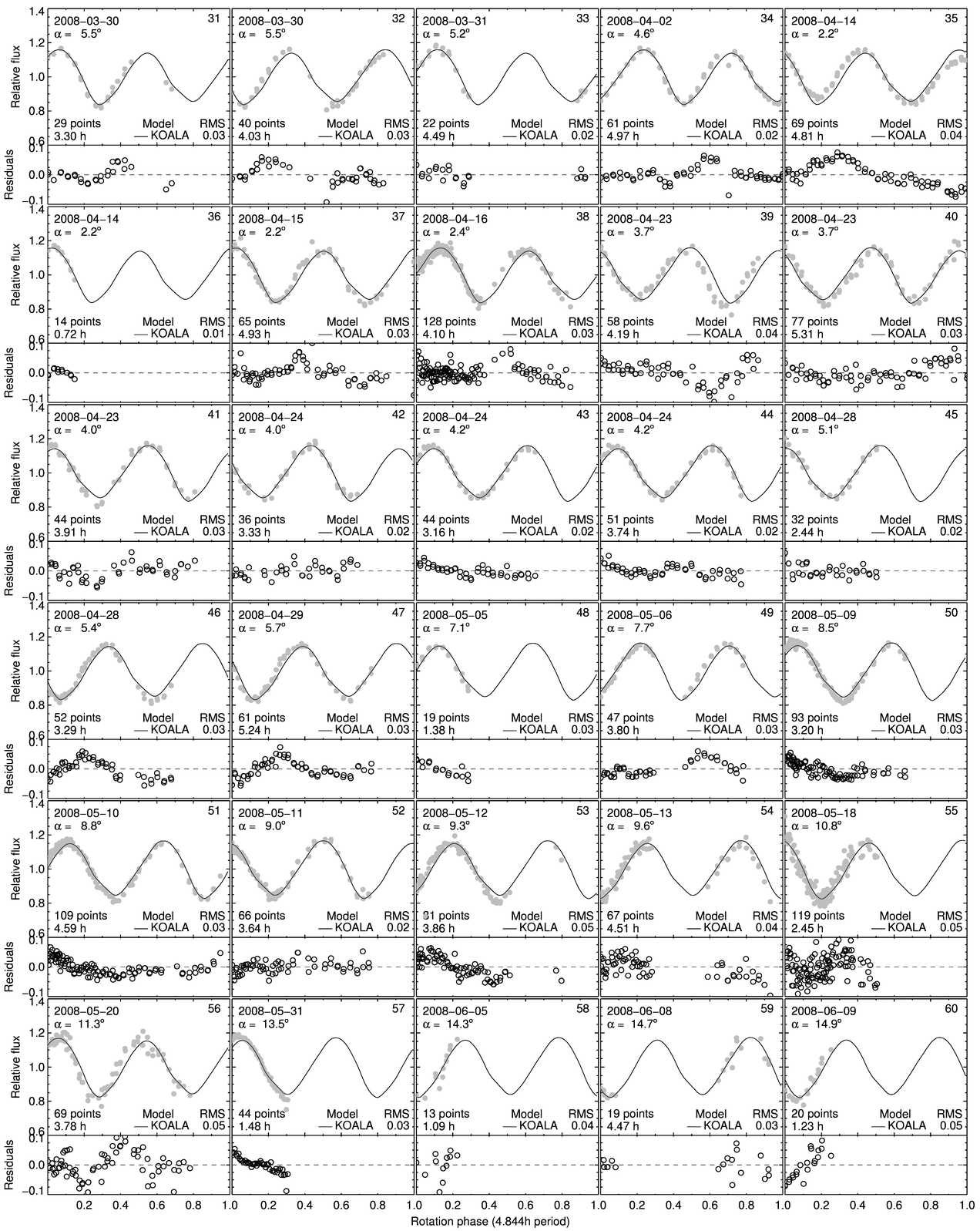}
  \end{subfigure}
  \caption{Suite \add{of all lightcurve plots, as described in Fig.~ \ref{fig:lc}.}}
\end{figure*}
\clearpage
\begin{figure*}[]
  \centering
  \ContinuedFloat
  \captionsetup{list=off}
  \begin{subfigure}{.9\textwidth}
    \includegraphics[width=\linewidth]{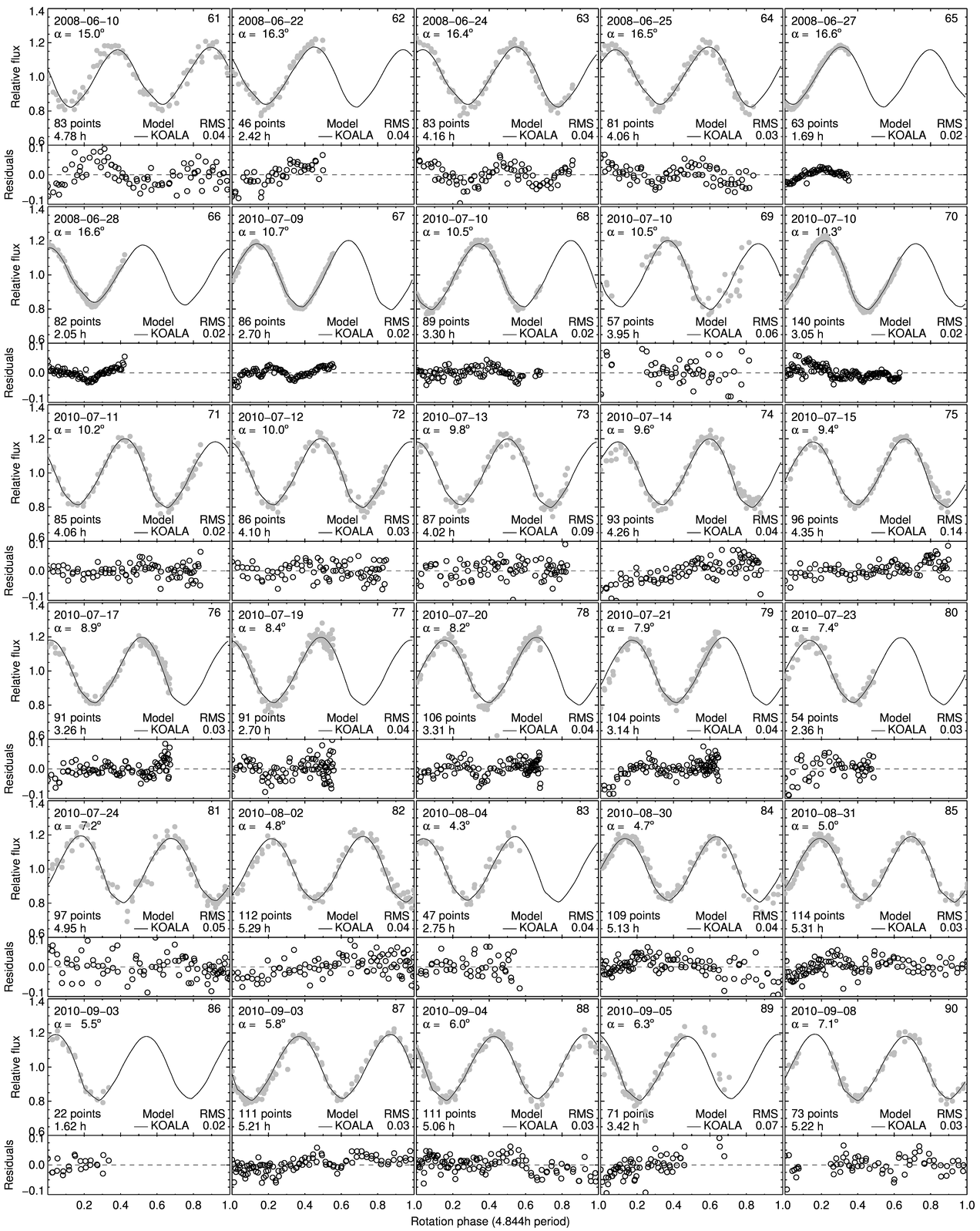}
  \end{subfigure}
  \caption{Suite \add{continued from previous page.}}
\end{figure*}
\clearpage
\begin{figure*}[]
  \centering
  \ContinuedFloat
  \captionsetup{list=off}
  \begin{subfigure}{.9\textwidth}
    \includegraphics[width=\linewidth]{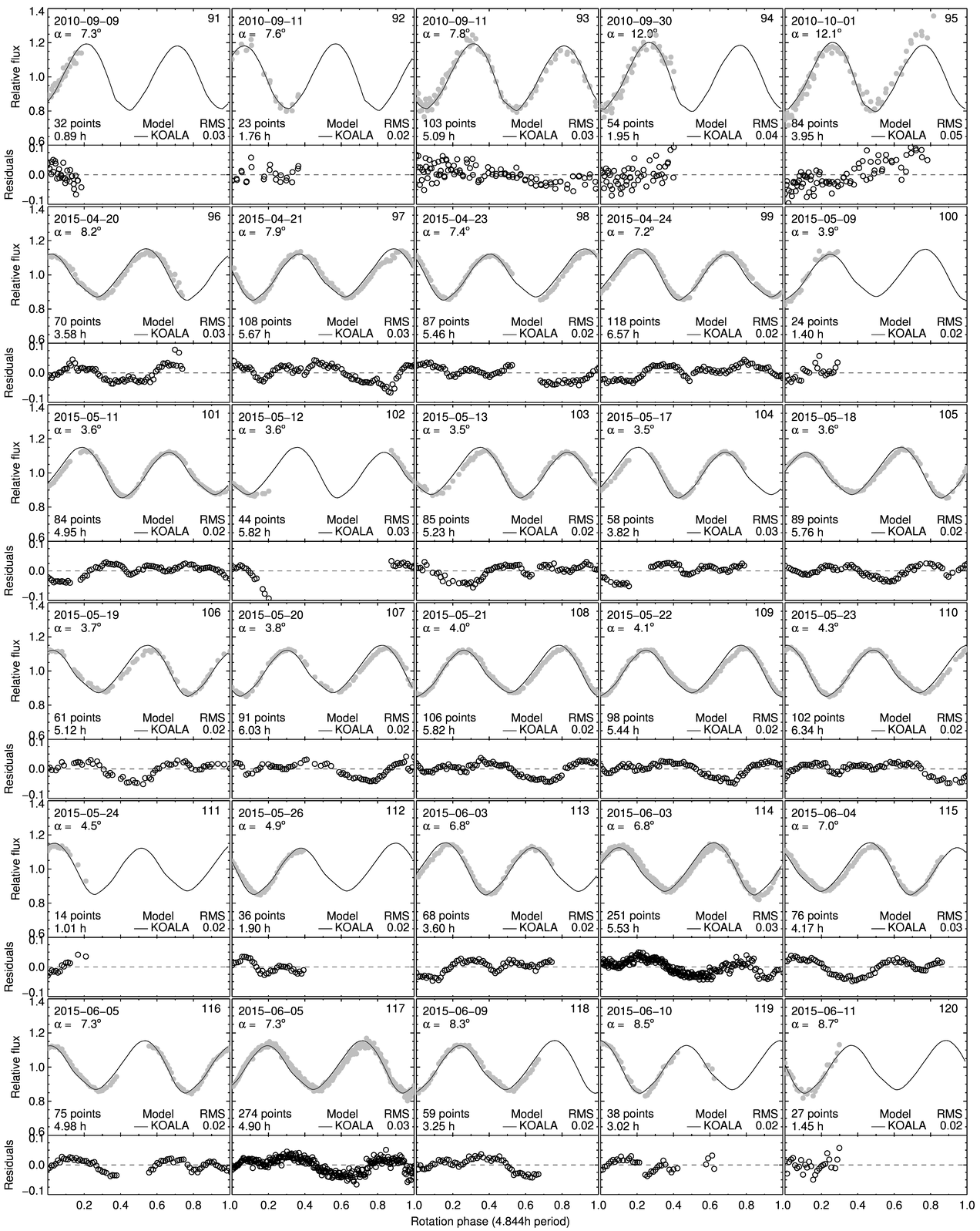}
  \end{subfigure}
  \caption{Suite \add{continued from previous page.}}
\end{figure*}
\clearpage
\begin{figure*}[]
  \centering
  \ContinuedFloat
  \captionsetup{list=off}
  \begin{subfigure}{.9\textwidth}
    \includegraphics[width=\linewidth]{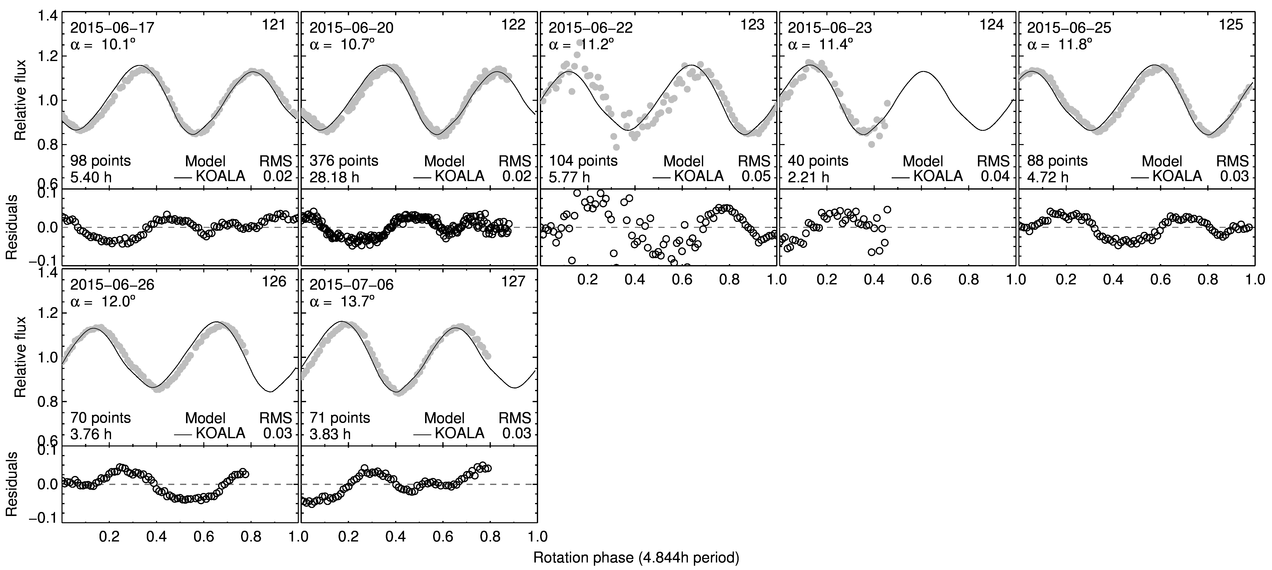}
  \end{subfigure}
  \caption{Suite \add{continued from previous page.}}
\end{figure*}

\clearpage

\section{Future occultations}

\begin{table*}
\begin{center}
  \caption{
    \add{Selection of stellar occultations by (107) Camilla scheduled for
    the next 3 years. For each, we report the 
    mean epoch of the event,
    the identifier of the UCAC-2 star and its magnitude ($m_\star$),
    the expected drop in magnitude ($\Delta m$), 
    the expected maximum duration of the event ($\Delta t$), 
    the uncertainty (3\,$\sigma$) on the position of
    both satellites \SatOne~and \SatTwo~at the date, projected on
    Earth, and the
    main area of visibility (location)}
    \label{tab:nextocc}
  }
  \begin{tabular}{ccrcrrrl}

	\hline \hline
     Mean epoch & 
     Star & 
     \multicolumn{1}{c}{m$_{*}$} & 
     \multicolumn{1}{c}{$\Delta m$} & 
     \multicolumn{1}{c}{$\Delta t$} & 
     \multicolumn{1}{c}{\SatOne} & 
     \multicolumn{1}{c}{\SatTwo} & 
     \multicolumn{1}{c}{Location}  \\
     (UTC) & (UCAC2) & \multicolumn{1}{c}{(mag)} &
     \multicolumn{1}{c}{(mag)} & (s) &
     \multicolumn{1}{c}{(km)} & \multicolumn{1}{c}{(km)} \\
   \hline

2018-06-05 21:25 & 3514 1714 & 13.6 & 2.8 &  5.6 & 90 & 1074 & Australia, Tasmania                    \\
2018-08-12 22:56 & 3605 4296 & 13.4 & 0.1 & 11.5 & 80 & 847  & Australia                              \\
2018-11-16 21:00 & 3404 4155 & 12.0 & 0.1 & 16.1 & 86 & 990  & South Africa, La R\'eunion Island (FR) \\
2018-12-13 02:38 & 3369 0629 & 12.4 & 0.6 & 22.5 & 93 & 841  & Chile, Argentina, Brazil (South)       \\
2020-01-04 17:16 & 3410 4468 & 12.2 & 0.1 & 23.4 & 79 & 896  & China, Japan                           \\
2020-01-21 13:53 & 3446 0788 & 11.9 & 0.1 & 17.1 & 88 & 880  & Australia, New Zealand (North)         \\
2020-02-10 13:47 & 3502 1656 & 11.9 & 3.1 & 17.2 & 94 & 771  & Australia, New Zealand (South)         \\
2020-02-13 23:50 & 3520 7286 & 12.0 & 0.1 & 17.9 & 96 & 1000 & Canada, Canary islands, Africa         \\

    \hline
   \end{tabular}
\end{center}
\end{table*}

\end{document}